\documentclass[fleqn,10pt]{wlscirep}
\usepackage[utf8]{inputenc}
\usepackage[T1]{fontenc}

\usepackage[numbers]{natbib}

\usepackage[utf8]{inputenc}
\usepackage{amsmath,amssymb}
\usepackage{siunitx} 
\usepackage{xspace}
\usepackage{multirow}
\usepackage{caption}
\usepackage{subcaption}
\usepackage{balance}
\usepackage{xr}
\usepackage{xstring}
\usepackage{lineno}
\usepackage{soul}
\usepackage{xcolor}

\newcommand{\rcmv}{\texttt{r/ChangeMyView}\xspace}
\newcommand{\R}[1]{\texttt{r/#1}\xspace}
\newcommand{\cmvdelta}{\ensuremath{\Delta}\xspace}
\newcommand{\noDelta}{\overbar{\Delta}}
\newcommand{\noDeltaSet}[1]{\overline{#1_{\Delta}}}
\newcommand{\overbar}[1]{\mkern 1.5mu\overline{\mkern-1.5mu#1\mkern-1.5mu}\mkern 1.5mu}

\begin{document}

\title{The language of opinion change on social media\\under the lens of communicative action}

\author[1]{Corrado Monti}
\author[2,3,*]{Luca Maria Aiello}
\author[1]{Gianmarco De Francisci Morales}
\author[1,4]{Francesco Bonchi}

\affil[1]{CENTAI, Torino, Italy}
\affil[2]{IT University of Copenhagen, Copenhagen, Denmark}
\affil[3]{Pioneer Centre for AI, Copenhagen, Denmark}
\affil[4]{Eurecat, Barcelona, Spain}

\keywords{Opinion change $|$ Social intent $|$ Communicative action $|$ Reddit}

\begin{abstract}
Which messages are more effective at inducing a change of opinion in the listener? We approach this question within the frame of Habermas' theory of communicative action, which posits that the \emph{illocutionary intent} of the message (its pragmatic meaning) is the key. Thanks to recent advances in natural language processing, we are able to operationalize this theory by extracting the latent social dimensions of a message, namely archetypes of social intent of language, that come from social exchange theory. We identify key ingredients to opinion change by looking at more than 46k posts and more than 3.5M comments on Reddit's \rcmv, a debate forum where people try to change each other's opinion and explicitly mark opinion-changing comments with a special flag called \emph{delta}. Comments that express no intent are about 77\% less likely to change the mind of the recipient, compared to comments that convey at least one social dimension. Among the various social dimensions, the ones that are most likely to produce an opinion change are knowledge, similarity, and trust, which resonates with Habermas' theory of communicative action. We also find other new important dimensions, such as appeals to power or empathetic expressions of support. Finally, in line with theories of constructive conflict, yet contrary to the popular characterization of conflict as the bane of modern social media, our findings show that voicing conflict in the context of a structured public debate can promote integration, especially when it is used to counter another conflictive stance. By leveraging recent advances in natural language processing, our work provides an empirical framework for Habermas' theory, finds concrete examples of its effects in the wild, and suggests its possible extension with a more faceted understanding of intent interpreted as social dimensions of language.
\end{abstract}

\flushbottom
\maketitle 

\section{Introduction}
\label{sec:intro}

The ``public sphere'', as theorized by J{\"u}rgen Habermas~\cite{habermas1962structural}, is the public arena wherein the democratic discourse develops. It plays a crucial role in the healthy functioning of a democracy,  as it allows citizens to shape public opinion, thus influencing policies and decisions~\cite{habermas1992further}. Such a role, today, is arguably filled in large part by the Internet and social media~\cite{gimmler2001deliberative,fuchs2014social}.

Habermas sees shared understanding, achieved through rational arguments, as the necessary pre-condition for social integration, and thus for democracy~\cite{habermas1981lifeworld}. The underlying assumption is that communication---and language in particular---is the only way to reach this shared understanding through grounded reasoning~\cite{habermas1979communication}. In particular, Habermas is interested in language from a formal pragmatics point of view, which differs from the socio-linguistic one. The focus is not on grammar, semantics, style, or sentiment, but rather on the interpretation of utterances~\cite{habermas1981reason}. The meaning of an utterance is not found in the sentence itself, as per the encoding/decoding paradigm of language, but in the intent of the speaker and in its reconstruction by the receiver, as per the intentionalist and dialogic paradigms~\cite{krauss1998language}. By loading language with intent, the speaker exercises an illocutionary force that can effectively change the hearer's mind~\cite{austin1962how} and eventually achieve cooperation based on a shared understanding of reality. This process, which Habermas calls \emph{communicative action}, can be triggered by potentially many different types of illocutionary forces~\cite{krauss1998language}, but especially by virtue of \emph{``shared knowledge, mutual trust, and accord with one another''}~\cite{habermas1979communication}.

The theory of communicative action is in stark contrast with how opinion dynamics has been traditionally operationalized.
Models of opinion change fail to take into account nuances of communication, let alone the intent of the actors involved, as they describe social interactions as one-dimensional events~\cite{grabisch2020survey}. These models have been necessarily oversimplified due to the complexity of quantifying social interactions. However, thanks to recent advances in natural language processing, we are now able to operationalize these concepts and measure the illocutionary force of an utterance, i.e., its intent. On one side, social theorists have identified universal hallmarks of communication that capture fundamental \emph{social dimensions} of pragmatics, namely archetypes of social intent that language can express (Table~\ref{table:dimensions}), on the other side computer scientists have developed methods to identify those dimensions from conversational text automatically and accurately~\cite{choi2020ten}. In this work, we employ these operationalizations to corroborate Habermas' hypothesis, and ask \emph{``which types of illocutionary intents are more effective at inducing a change of opinion in the listener?''}.

We seek to answer this question in the wild. While there have been some attempts in a similar direction~\cite{Rosenberg2005}, our approach is general and widely applicable, employs automated coding (instead of a manual one), and makes use of large-scale data, obtained from a public discussion forum. In particular, we analyze data from Reddit's \rcmv, an on-line forum where people debate positions and try to change each other's opinion and reach some form of consensus: a success of the communicative action. This forum is particularly suited for our purposes as it provides a ground truth of opinion change. The debaters, in fact, explicitly recognize convincing arguments that changes their mind, thus producing a sort of consensus. Moreover, people participating in the forum have an epistemic goal~\citep{habermas2018interview}, as they are asked to approach the discussion `in an effort to understand other perspectives on the issue' and `with a mindset for conversation' (\url{https://www.reddit.com/r/changemyview}). Therefore, the forum presents close to an ideal Habermasian communication situation~\citep{habermas2018interview}, whereby perlocutionary acts and strategic actions~\cite{habermas1979communication} have no reason to be~\citep{heng2003habermas}. The pragmatic illocutionary intent, i.e., the social dimensions we employ, are therefore representations of the communicative action happening among participants of the forum.

Our results align with the hypothesis of Habermas: the three most important social dimensions for a convincing argument are the conveying of \emph{knowledge}, the appeal to \emph{similarity}, and the expression of \emph{trust}. Conversely, messages that do not clearly convey a social intent are exceedingly unlikely to change someone's view ($77\%$ less, as illustrated in Figure~\ref{fig:odd-ratios}). In addition, we identify particular intents that are more or less effective at changing opinions when replying to a specific expressed intent. For instance, responding to posts containing tones of conflict by signaling group identity is less effective at changing the mind of the poster. Conversely, we find that expressing conflict actually improves the odds of changing someone's mind, especially when replying to an already conflictive stance.

While Habermas' theory is vast and far-reaching, in this study we focus on the aspects related to the pragmatic intent of language, and its effects on opinion change.
Our contribution lies in the operationalization of this aspect through the lens of social dimensions of pragmatics inspired by social exchange theory~\citep{blau64exchange}.
We find empirical support to the theory by looking at text in the wild from a popular online discussion forum, and by using fully-automated means of coding.

\section{Research design}
\label{sec:research-design}

\subsubsection*{Gathering public sphere discussions from Reddit}
\label{sec:selecting-sociopol}

Reddit is an online forum organized in topical communities, called \emph{subreddits}. Inside a subreddit, users can publish \emph{posts}, or \emph{comment} in response to other posts or comments. In the \rcmv subreddit, posters express a point of view that commenters attempt to change. Comments that succeed in doing so receive from the poster a token of merit called \emph{delta} (\cmvdelta) to symbolize a successful attempt at opinion change. We limit the scope of our study to \rcmv posts dealing with sociopolitical issues. This way, we bring the object of our study closer to the public sphere discussion as conceptualized by Habermas, while also ensuring a higher topical homogeneity. Following the operative definition given by Moy and Gastil~\cite{moy2006predicting}, we define a post as \emph{sociopolitical} if it is about at least one of the following categories:
  (i)~political figures, parties or institutions;
  (ii)~broad cultural and social issues (e.g., civil rights, moral values);
  (iii)~national issues (e.g., healthcare, welfare).

To automatically categorize \rcmv posts as sociopolitical or not, we train a supervised classifier on Reddit posts (details in Section~\ref{sec:sociopolitical-classifier}). Out of the \num{65727} \rcmv posts with textual content, we identify \num{46046} as sociopolitical: \num{20239} of these have at least one comment with \cmvdelta ($P_{\Delta}$), whereas \num{25807} posts do not have any~($\noDeltaSet{P}$). Those \num{46046} sociopolitical posts received \num{3690687} comments, which we split in two sets: one containing the \num{38165} comments that received a \cmvdelta ($C_{\Delta}$) and one containing the remaining comments ($\noDeltaSet{C}$). Summary statistics about the Reddit dataset are provided in Figure~SI1.

\subsubsection*{Capturing social intent from language}
\label{sec:design:10dims}

To infer the social intent that Reddit messages convey, we ground our analysis in a theoretical model of \emph{social dimensions} that reflect fundamental social aspects of the pragmatics of language (Table~\ref{table:dimensions}). In ordinary conversations, these dimensions are often verbalized to signal social intent, for example to confer appreciation or to give emotional support. These dimensions have been identified through an extensive survey of social science research~\citep{deri2018coloring} and they are comprehensive of some of the most influential categorizations of social interactions~\citep{wellman1990different,fiske1992four,spencer2018rethinking}. Thus, we use our analysis as a test bed for a novel natural language processing model~\citep{choi2020ten}, able to capture with high accuracy expressions of the social dimensions in conversational language (details in Section~\ref{sec:methods:10dims}). Given an input message $m$ and a social dimension $d$ from Table~\ref{table:dimensions}, the tool produces a score $s_d(m)$ that represents the likelihood that message $m$ contains social dimension $d$. To ease the interpretation of the results, we binarize the scores to split messages between those that carry dimension $d$ with high probability and those that do not (see Section~\ref{sec:normalization-scores}).

Our study relies on estimating the effect of the intent extracted from Reddit comments on the behavior of the user who reads them. As such, it can be sorted under the `text-as-treatment' umbrella within the causal inference literature~\cite{feder2021causal}. Feder et al.~\cite{feder2021causal} specifically identify three requirements for proper causal inference under the potential outcomes framework: \emph{ignorability}, \emph{positivity}, and \emph{consistency}. The \emph{ignorability} assumption requires the treatment assignment to be independent of the realized counterfactual outcomes. In our case, while the treatment is not randomly assigned, the author of the post does not have control on who writes an answer. This fact ensures that there is no selection bias due to the choices of the poster. The focus on a narrow topic reduces the possibility of unobserved confounders (e.g., a specific social dimension is more present on a topic for which it is easier to change opinion). The interaction between users happens exclusively via the text, so it is unlikely that there are other unobserved confounders that have effect both on the social dimensions and on the opinion of the original author of the post (e.g., body posture or voice pitch). One possible source of confounding is the profile of the poster which contains their posting history, which, if viewed, might influence the other users, and naturally influences the opinion of the original author.
While we cannot completely exclude this confounder, judging a poster by anything other than their argument goes against the spirit of \rcmv, thus we assume that this possible confounder plays a negligible role. \emph{Positivity} is the assumption that the probability of receiving treatment is strictly bounded between $0$ and $1$. This assumption is easily verified empirically in our case. Finally, \emph{consistency} requires that the observed outcome at a given treatment status for an individual is the same as would be observed if that individual was assigned to the treatment. In practice, for the purpose of assuming consistency and making valid inferences, it is necessary to develop the measure of the treatment with different data than the data used to estimate the causal effect, such that there is no interference.
In our case, the measure of social intent we use has been developed separately~\citep{choi2020ten} and is not related to the opinion change outcome.

\begin{table}[t]
\centering
\footnotesize
\caption{Social dimensions of relationships historically identified in social sciences and surveyed by Deri at al.~\cite{deri2018coloring}.}
\label{table:dimensions}
\vspace{-1mm}
\begin{tabular}{ll}
\toprule
\textbf{Dimension} & \textbf{Description}  \\
\midrule
\emph{Knowledge} & Exchange of ideas or information; learning, teaching~\cite{fiske2007universal}  \\
\emph{Power} & Having power over behavior and outcomes of another~\cite{blau64exchange} \\
\emph{Status} & Conferring status, appreciation, gratitude, or admiration~\cite{blau64exchange}  \\
\emph{Trust} & Will of relying on the actions or judgments of another~\cite{luhmann1982trust}  \\
\emph{Support} & Giving emotional or practical aid and companionship~\cite{fiske2007universal}  \\
\emph{Similarity} & Shared interests, motivations or outlooks~\cite{mcpherson2001birds} \\
\emph{Identity} & Shared sense of belonging to the same group~\cite{tajfel2010social} \\
\emph{Fun} & Experiencing leisure, laughter, and joy~\cite{argyle2013psychology}  \\
\emph{Conflict} & Contrast or diverging views~\cite{tajfel1979integrative} \\
\bottomrule
\end{tabular}
\vspace{-3mm}
\end{table}

\subsubsection*{Quantifying communicative action}
\label{sec:methods:delta_nodelta}

To assess the interplay between social intent and opinion change, we define suitable probabilities and odds ratios which are detailed in Section~\ref{sec:odds-ratios-definitions}.
We consider the odds ratios (OR) of finding dimension $d_i$ in comments with a \cmvdelta (or posts that awarded a \cmvdelta), compared to those without \cmvdelta.
To study the interactions between posts and comments, we need a more elaborate model to account for their complex relationship, which we base on the dialogic interpretation of language (see Section~\ref{sec:odds-ratios-definitions}).
We therefore compute the increase of the probability---compared to random chance---that a comment with dimension $d_i$ would receive a \cmvdelta given that its corresponding post contains dimension $d_j$.

To account for possible confounders and to assess the significance of our findings, we complement the analysis of probabilities with logistic regression models. In particular, we consider the original ideological leaning of the two involved individuals because it is a factor that several opinion models take into account~\cite{grabisch2020survey}. As a proxy for such leaning, we extract different sets of variables from individuals' participation in partisan sociopolitical groups~\cite{massachs2020roots}. We then use this information to test three alternative sets of confounders. First, \emph{individual political sides}: whether each of the two involved individuals participates to a left- or right-wing group. Second, \emph{interaction of political sides}: whether the two individuals both participate to polarized groups, and whether they are on opposing sides. Third, we check \emph{shared groups}, i.e., whether the two individuals participate in exactly the same political and polarized group. Finally, to make sure that the signal captured by the social dimensions is not mere sentiment polarity, we add positive and negative sentiment of the message as controls. The considered confounders are summarized in Table~\ref{tab:homophily-log-reg-explaination} and
further detailed in Section~\ref{sec:confounders}.

\section{Results}
\label{sec:results}

\begin{figure*}[t]
    \centering
    \begin{tabular}{ccc}
      % (a) Posts & (b) Comments & (c) Interactions \\
      \raisebox{-0.43\height}{\includegraphics[width=0.25\linewidth]{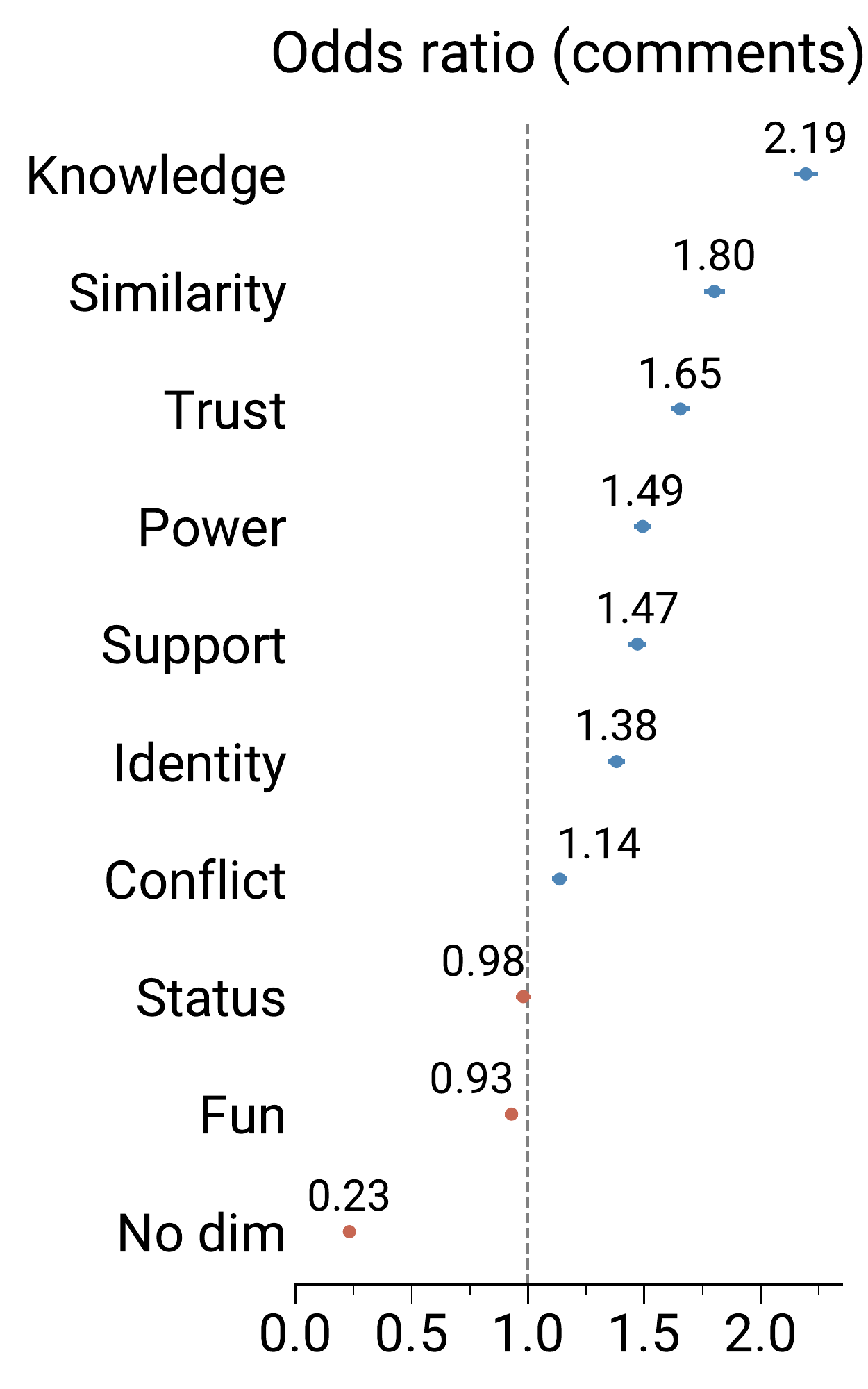}} &
      \raisebox{-0.43\height}{\includegraphics[width=0.25\linewidth]{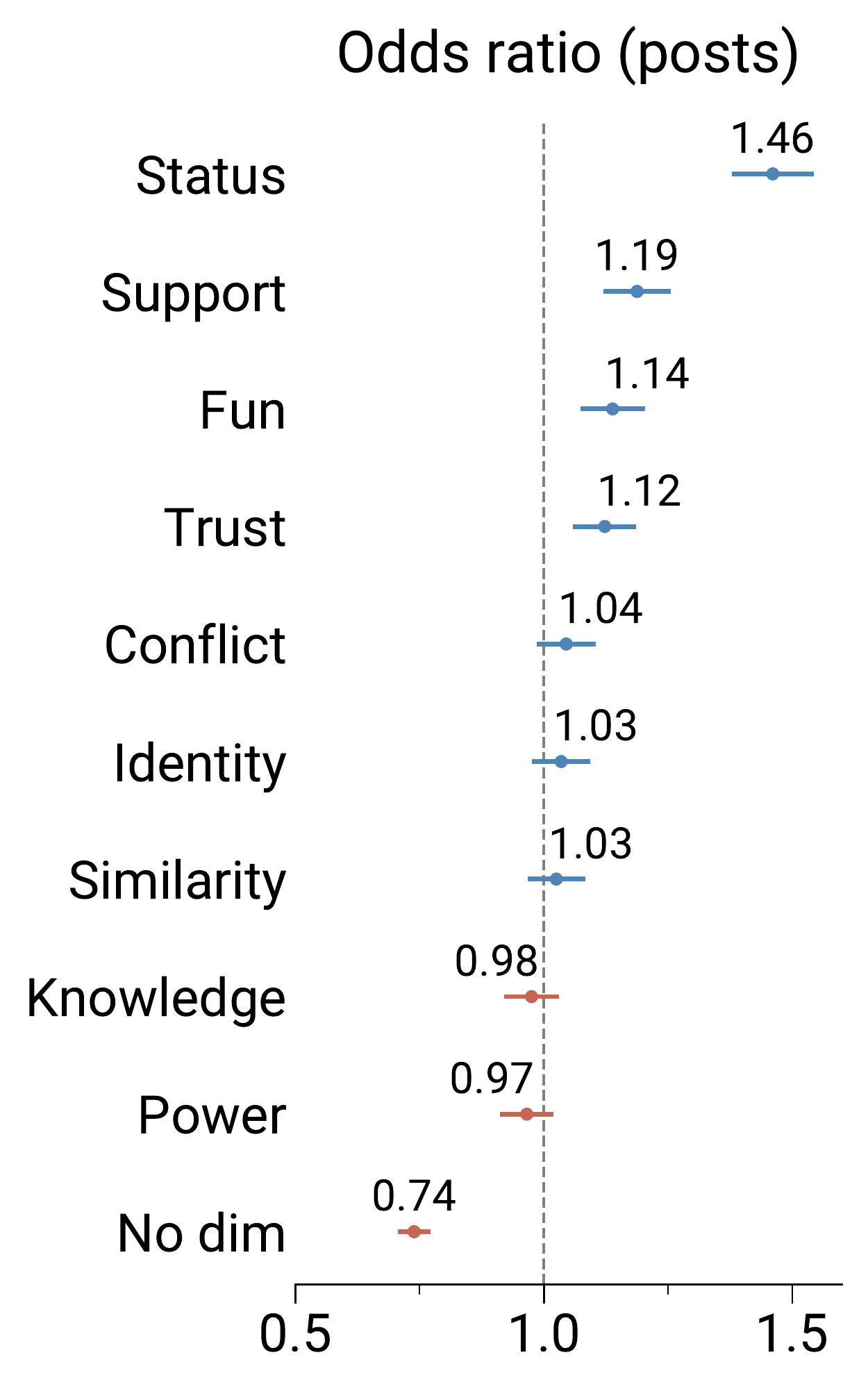}} &
      \raisebox{-0.4\height}{\includegraphics[width=0.45\linewidth]{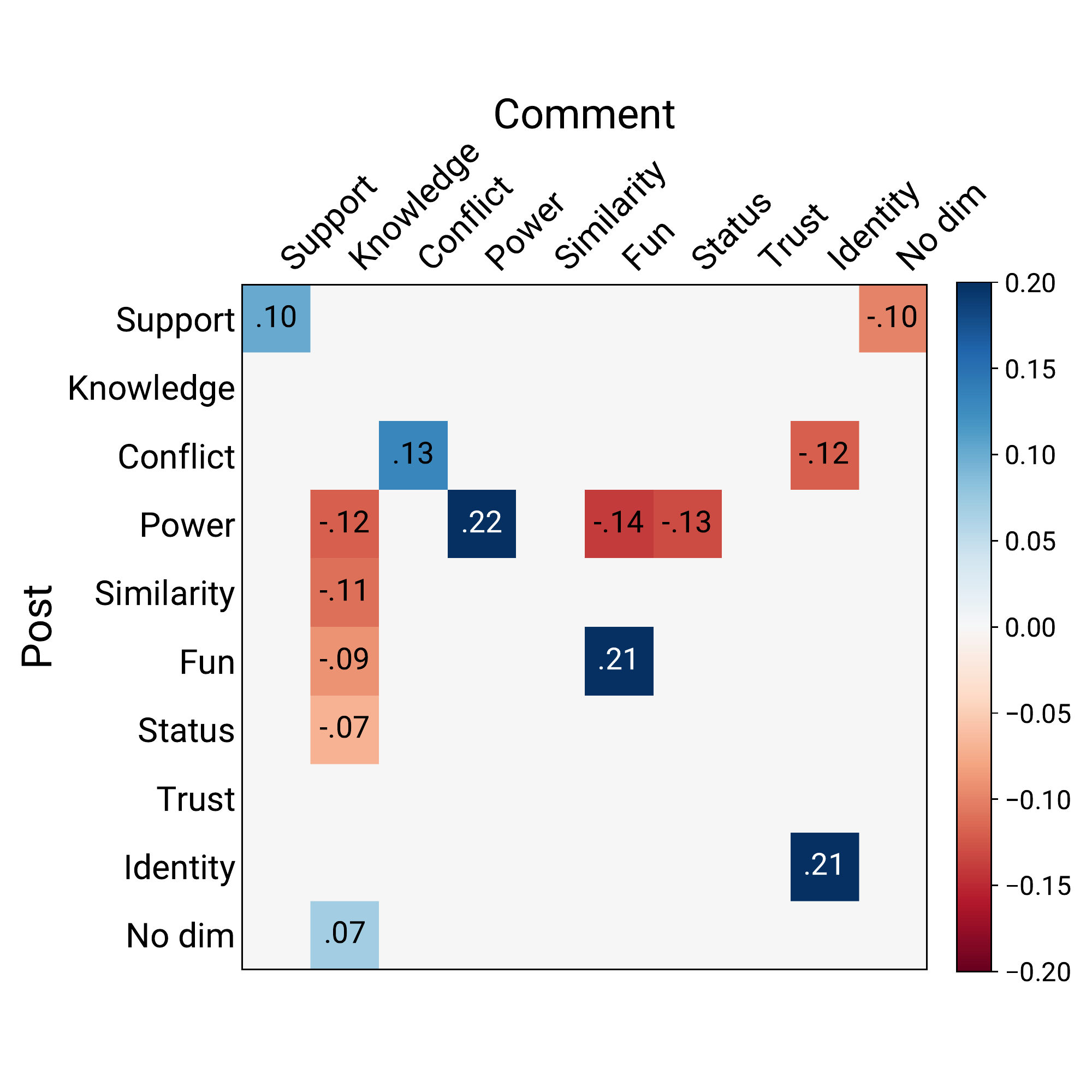}} \\
      (a) & (b) & (c) \\
    \end{tabular}
    \caption{Odd ratios of containing a dimension (a) in comments that were successful in changing the poster's opinion versus those that were not, and (b) in posts expressing opinions that were changed by other community members versus posts that did not experience any opinion change. Error bars represent 95\% confidence intervals. On the right (c), we report only the statistically significant odds ratios ($p < 0.01$) for interactions between dimensions in comments and posts. Cells represent the variation of the probability of achieving a \cmvdelta given a combination of dimensions.
   \label{fig:odd-ratios}
   }
\end{figure*}

We design our study to answer three research questions:

\vspace{2pt}\noindent \textbf{RQ1.} Are messages that convey a social dimension more likely to change the opinion of the reader?

\vspace{2pt}\noindent \textbf{RQ2.} Which \emph{types} of social dimension are more often present in opinion-changing messages?

\vspace{2pt}\noindent \textbf{RQ3.} Which \emph{combination} of intents expressed by the poster and the commenter are more likely to result in an opinion change?

\vspace{3pt}

To find whether expressions of social intent matter in the process of opinion change (RQ1), we compute the odds ratios $OR_{C_{\Delta}}(d_i)$ of a social dimension $d_i$ being conveyed by comments with \cmvdelta, compared to comments with no \cmvdelta (Figure~\ref{fig:odd-ratios}a). Comments that express no intent exhibit an odds ratio of 0.23, meaning that they are about 77\% less likely to change the mind of the recipient, compared to comments that convey at least one social dimension. To confirm the significance of this result, we apply a logistic regression model by using all the dimensions as independent variables, and whether the comment received a \cmvdelta as dependent variable.

We find (Table~SI4) that all dimensions have a significant association with the message being considered as view-changing, with the notable exception of \emph{fun}, that we therefore remove from further analysis. All the other dimensions emerge as significant. In general, social intents are positively associated with opinion change, except for \emph{status}, that exhibits a significant yet negative relationship instead. Status captures admiration, appreciation, and praise. Because these expressions are typically more associated with agreement than with disagreement, it is likely that comments conveying status are meant to support the point of view of the original poster rather than attempting to change it. Thus, these comments are less likely to receive a $\Delta$.

To check the robustness of the association between social dimensions and $\Delta$ to the inclusion of other factors, we fit regression models that use the sets of confounders we presented (Table~\ref{tab:homophily-log-reg}). The confounders do not change the main result, and social dimensions keep consistently emerging as significant: for instance, in model $E$ we observe that the inclusion of sentiment scores do not alter the significance of social dimensions. This is true even if we consider also the length of the message (Table~SI5). Moreover, for each set of confounders, we test what happens with and without considering the social dimensions. Beside statistical significance, we assess quality of fit as measured by the adjusted Pseudo-$R^2$ metric. All the models that use social dimensions (models~\mbox{$E$-$H$}) have a higher quality of fit than those that only use the ideological positions of the authors (models~\mbox{$A$-$D$}). In particular, the best model that does not consider social dimensions but only sentiment and political group (model~$D$) has a quality of fit that is roughly half of the model when with social dimensions (model~$H$). The results presented in Table~\ref{tab:homophily-log-reg} are robust to data imbalance (Table~SI6); also, the significance of the coefficients is not caused by random fluctuations in the data: a randomized regression model with reshuffled variables across examples loses all statistical significance (Table~SI7).

In summary, messages that convey a social intent are more likely to be associated with opinion change in the intended reader, even after controlling for confounding factors. This finding backs our main hypothesis that considering social dimensions is essential to correctly model opinion change.

To assess what types of social intent are associated with comments that are successful in changing people's opinion (RQ2), we compare the magnitude of the odds ratios across the different social dimensions (Figure~\ref{fig:odd-ratios}a). Comments that received a \cmvdelta are exceedingly more likely to convey \emph{knowledge} than those with no \cmvdelta  ($+119\%$). To find illustrative examples of argumentative nuances associated with comments conveying different dimensions, we inspected manually the messages with $\Delta$ that were classified with high confidence. We report these examples next.

The classifier assigns high knowledge scores to messages that provide logical reasoning (e.g., \emph{``The NHS isn't free, it's free at the point of use, we pay for it through tax contributions''}), refer to factual evidence (\emph{``Door levers are more likely to fail over time since they require springs.''}), cite sources (\emph{``History shows us that the benefits of widespread vaccinations greatly outweigh the risks of possible resistant mutations''}), and present points of view that might be debatable yet stem from factual observations of the world (\emph{``The automobile market is oversaturated with cars that run solely on gasoline''}).

\begin{table*}[t]
    \centering
    \scriptsize
    \caption{ %
    Odds ratios obtained by logistic regression.
    Each column corresponds to a model with a specific set of variables.
    A description of each confounder is given in Table~\ref{tab:homophily-log-reg-explaination}.
    We indicate with asterisks the statistically significant correlations (with one, two or three asterisks corresponding to $p < 0.05$, $p < 0.01$ and $p < 0.001$ respectively).
    P-values are corrected according to the Benjamini-Hochberg procedure~\cite{benjamini1995controlling}, to reduce the chance of spurious correlation emerging because of the high number of factors we consider.
    }
    \label{tab:homophily-log-reg}

\begin{tabular}{lllllllll}
\toprule
{} &         A &         B &         C &         D &         E &         F &         G &         H \\
\midrule
Adj. Pseudo-$R^2$            &   0.00305 &   0.00446 &   0.00720 &   0.00935 &   0.01643 &   0.01780 &   0.02050 &   0.02265 \\ \midrule \midrule
Intercept                    &  0.010*** &  0.011*** &  0.011*** &  0.011*** &  0.010*** &  0.010*** &  0.010*** &  0.010*** \\ \addlinespace
Support                      &  \dotfill &  \dotfill &  \dotfill &  \dotfill &  1.096*** &  1.095*** &  1.097*** &  1.098*** \\
Knowledge                    &  \dotfill &  \dotfill &  \dotfill &  \dotfill &  1.217*** &  1.216*** &  1.216*** &  1.216*** \\
Conflict                     &  \dotfill &  \dotfill &  \dotfill &  \dotfill &  1.024*** &  1.026*** &  1.025*** &  1.024*** \\
Power                        &  \dotfill &  \dotfill &  \dotfill &  \dotfill &  1.097*** &  1.099*** &  1.097*** &  1.096*** \\
Similarity                   &  \dotfill &  \dotfill &  \dotfill &  \dotfill &  1.110*** &  1.108*** &  1.110*** &  1.112*** \\
Status                       &  \dotfill &  \dotfill &  \dotfill &  \dotfill &  0.930*** &  0.929*** &  0.934*** &  0.937*** \\
Trust                        &  \dotfill &  \dotfill &  \dotfill &  \dotfill &  1.143*** &  1.143*** &  1.144*** &  1.144*** \\
Identity                     &  \dotfill &  \dotfill &  \dotfill &  \dotfill &  1.085*** &  1.086*** &  1.086*** &  1.086*** \\ \addlinespace
Sentiment Pos.               &  1.174*** &  1.173*** &  1.175*** &  1.177*** &  1.110*** &  1.110*** &  1.110*** &  1.112*** \\
Sentiment Neg.               &  1.080*** &  1.082*** &  1.081*** &  1.081*** &  1.056*** &  1.058*** &  1.058*** &  1.057*** \\ \addlinespace
Comment Left-Wing            &  \dotfill &     1.014 &  \dotfill &  \dotfill &  \dotfill &     1.005 &  \dotfill &  \dotfill \\
Comment Right-Wing           &  \dotfill &  0.790*** &  \dotfill &  \dotfill &  \dotfill &  0.795*** &  \dotfill &  \dotfill \\
Post Left-Wing               &  \dotfill &  0.867*** &  \dotfill &  \dotfill &  \dotfill &  0.873*** &  \dotfill &  \dotfill \\
Post Right-Wing              &  \dotfill &  0.852*** &  \dotfill &  \dotfill &  \dotfill &  0.849*** &  \dotfill &  \dotfill \\ \addlinespace
Both Polarized               &  \dotfill &  \dotfill &  0.321*** &  \dotfill &  \dotfill &  \dotfill &  0.324*** &  \dotfill \\
Both Polarized \& Diff. Side &  \dotfill &  \dotfill &  2.555*** &  \dotfill &  \dotfill &  \dotfill &  2.510*** &  \dotfill \\
Diff. Side                   &  \dotfill &  \dotfill &     1.022 &  \dotfill &  \dotfill &  \dotfill &     1.023 &  \dotfill \\
Shared Group                 &  \dotfill &  \dotfill &  \dotfill &  0.286*** &  \dotfill &  \dotfill &  \dotfill &  0.287*** \\
\bottomrule
\end{tabular}

\end{table*}

Successful comments are $80\%$ more likely to allude at \emph{similarity} between the stance of the poster and the commenter (\emph{``I'm glad to know we agree on this''}) or between their experiences (\emph{``My friends used to live in a large city in Asia too''}), and $65\%$ more likely to contain language that discloses \emph{trust} towards entities relevant to their argument (\emph{``I believe what they're saying''}, \emph{``The people causing problems are a tiny minority compared to the reasonable majority''}). Appeals to power (\emph{``The only rights which exist in objective reality are legal rights''}), expressions of support (\emph{``I can understand being disappointed, but ...''}, \emph{``I'd feel sympathy for their situation''}), and language markers of group identity (\emph{``They are members of their tribe, they don't necessarily want to be part of the larger nation.''}) are also more frequent in comments that received a \cmvdelta.
These results are corroborated by the coefficients of the regression models. In all of our models (Table~\ref{tab:homophily-log-reg}) the coefficients obtained by each social dimension are extremely stable: each odds ratio varies by less than $0.01$.

In summary, our results provide empirical evidence for the presence of the \emph{``shared knowledge, mutual trust, and accord with one another''} that Habermas identified as the founding pillars of communicative action. Indeed, the three social intents of \emph{knowledge}, \emph{trust}, and \emph{similarity} characterize messages with a \cmvdelta more than all the other intents.

Messages written in an attempt to change someone's opinion are not isolated entities; rather, they are part of an on-going conversation between two parties. We study the simplest form of such interactions in Reddit, namely the relationship between the intent expressed by the post which starts the conversation and that expressed by the aspiring view-changing comment (RQ3). Interestingly, posts by people who change their view are characterized by different social dimensions from those found in view-changing comments (Figure~\ref{fig:odd-ratios}b). Most prominently, conversation-starting messages written by individuals who end up awarding a \cmvdelta are $46\%$ more likely to convey \emph{status}---words of appreciation or gratitude. The presence of status is an indication of approaching the dialogue with respect, either towards the subject of the discussion (\emph{``I have nothing but the utmost respect for service men and women, but ...''}) or the discussion itself (\emph{``I'd really appreciate if someone could help me quantify and qualify my views''}). Conversely---and in agreement with the observation that partisans abiding to power have a less objective view of reality~\cite{keltner1997defending}---people who introduce their opinion by appealing to power or mentioning power dynamics (\emph{``If people were required to vote, they would take more of an interest in the political situation''}) are the least likely to grant a \cmvdelta.

Discussions that lead to a change in opinion are often those where the intent of the poster receives a response motivated by a similar intent. Figure~\ref{fig:odd-ratios}c shows a matrix of interaction between the intent of the poster and that of the commenter. Cells represent the variation of the probability of achieving a \cmvdelta given a specific combination of intents. Comments that receive a \cmvdelta are more likely to express an intent that matches that of the original poster, at least for five out of all the social dimensions that our tool captures. For example, when a post intends to convey a \emph{power} dynamic, the most effective response is to make a similar appeal to power ($+22\%$ more likely of getting a \cmvdelta). This observation is in line with the general principle of \emph{reciprocity}~\cite{gouldner1960norm}, and with the interpretation of conversations as social exchanges that occur under the assumption that a contribution of a certain type should be matched by a response of a similar type~\cite{blau64exchange}.

Some combinations of dimensions that break this symmetry are less likely to reach an agreement. When the poster expresses \emph{power}, comments replying with \emph{status} are $14\%$ less likely to receive a \cmvdelta. Power-status dynamics occur frequently in social relationships. Individuals entertaining relationships with people who have power over them tend to maintain those relationships stable by means of providing status back~\cite{blau64exchange} (e.g., employees expressing admiration for their manager). Contrary to typical social norms, the aim of \rcmv is not to maintain stability, but rather to disrupt ideas, a process that is not well-supported by power-status dynamics. Knowledge and fun are also less effective responses to power ($-12\%$ and $-13\%$ respectively).

Responding with comments containing markers of group identity is less effective in changing the mind of the poster, if the discussion was initiated with tones of conflict ($-12\%$).
Identity and conflict are tightly coupled in human societies.
When brought up in debates, expressing identity often sparks conflict with those who do not feel (or are not entitled to) the same sense of belonging. Symmetrically, identity is also a typical way to oppose conflict, as it is signaled as a way to manifest in-group defense in response to an out-group aggression~\cite{tajfel1979integrative}. Such dynamics are more likely to originate disagreement than to facilitate convergence of ideas.

Last, despite knowledge-based reasoning being the most effective approach to changing someone's view (Figure~\ref{fig:odd-ratios}a), its effectiveness is highest when responding to posts that are not strongly characterized by a social intent, such as those that plainly state an opinion (e.g., \emph{``I believe that the individual is not more vital then the global scale.''}). When responding to posts that are strongly characterized by an intent (especially those expressing power), knowledge-conveying comments are less effective. Even though our tools of analysis cannot ascertain the root cause of this phenomenon, we speculate that posts conveying a clear intent might indicate stronger motivated reasoning, or stronger ideological biases of the poster, both of which are shields to persuasion.

\section{Discussion}
\label{sec:discussion}

We have shown that the pragmatic intent of the dialogue between two parties has an important effect on the success of communicative action, as theorized by Habermas, and in reaching agreement.
In particular, the social dimensions that are most indicative of a successful communicative action are the ones originally indicated by Habermas: knowledge, similarity, and trust. Results show robustness across different settings (see Figure~SI5 and Table~SI5). These findings echo the interpretation of Habermas' theory provided by Terry~\citep{terry1997habermas}: in an ideal speech situation \emph{``all participants must [...] refer to facts and knowledge with which all are familiar, contribute to the discussion in an open, honest way, and be prepared to place themselves in the position of others in order to understand the latter's point of view''}.
However, other dimensions that have not previously been identified as relevant also have a positive effect, such as appeals to power or empathetic expressions of support. In line with theories of constructive conflict~\cite{follett2011constructive}, yet contrary to the popular characterization of conflict as the bane of modern social media, we found that in the context of a structured public debate, voicing conflict can promote integration, especially when it is used to counter a conflictive stance.
We find indeed that belonging to different ideological sides increases the chances of one of the parts to change their opinion (Table~\ref{tab:homophily-log-reg}, models $G$ and $H$). This finding is further detailed in Figure~SI6. Finally, we find that sentiment, as commonly measured in natural language processing, by itself is an unreliable indicator for opinion change. Indeed, the definition of sentiment commonly used (`the underlying feeling, attitude, evaluation, or emotion associated with an opinion'~\citep{cambria2017practical}) is too broad for our purposes.

These results, combined, point towards an extension of the original Habermasian theory which includes a more faceted understanding of intent, interpreted as social dimensions of language. In particular, the original theory of communicative action is quite broad in scope, but only provides a few concrete examples. This study contributes to materialize it, by providing an empirical descriptive framework for it, and by finding concrete examples of its effects in the wild. In fact, while Habermas' theory aims at a high level of generality and abstraction, able to unify different levels of the theory of communication, it still places value on its ability to receive empirical confirmations. As noted by Bohman~\cite{bohman1999theories}, practical verification of the consequences of such theoretical constructs should be the main route to solve the problem, posed by pluralism, of choosing between alternative theories. Thanks to the explosive growth of available data and of our ability to process it, we are able to connect such large and general theories to empirical and falsifiable claims~\cite{buyalskaya2021golden}. We focus on one specific aspect of the more general theory of communicative action; specifically, on understanding which types of illocutionary intents are more effective at reaching consensus.

Other works have previously explored how to operationalize habermasian ideas by measuring the nature of discourse, typically in the study of deliberative democracies~\cite{steiner2005deliberative}. However, these works present two profound differences with our approach. First, we focus on how the pragmatic intent of language affect communication between social media users, rather than communication performed in controlled settings or by professional politicians~\cite{steenbergen2003measuring}. Second, in order to achieve this goal on the large scale of social media data, we apply machine learning techniques to automatically classify intents, rather than relying on human evaluators as previously done in the literature~\cite{gerber2014deliberative}. We leverage such data availability of the Web in order to operationalize the study of communication as it happens in the wild, studying how social media contribute to the public sphere. In this sense, we follow Habermas' idea of communication operating with similar mechanisms at different levels.

The link between social exchange theory~\cite{blau64exchange}, on which the social dimensions are based, and the theory of communicative action, has been missing from the pragmatics literature, but is clearly worth exploring further in light of the results presented here. Social exchange theory is focused  explaining an equilibrium in social relationships, and provides an economics-based framework for understanding social behavior. Conversely, the theory of communicative action explains (among other things) a change---i.e., reaching a consensus---and insists that strategic actions based on cost-benefit analysis are counter to this ultimate goal. Our study, based on social media data, provides empirical support for both theories, and yet hints at a larger picture: a theory that is a synthesis of both sources while resolving the tension existing due to their different goals. Creating such a theory is a challenging endeavor, but also an exciting one given its potential broad impact. Our study is but a first step in this direction that shows a working, proof-of-concept operationalization of social dimensions extracted from language, and its importance in understanding the consensus-reaching process. Our work can help guiding the design of communication campaigns (e.g., against vaccine hesitancy) by better understanding the potential effectiveness of alternative strategies through the analysis of their social intent.

Future work could improve on our analysis in five main aspects. First, our social dimensions model alone is not predictive of whether a comment will get a \cmvdelta, as this is not the focus of our work. Such prediction is hard to make not only because of sparsity (successful comments are overwhelmingly outnumbered by unsuccessful ones), but mainly because of the intrinsic limit to predictability of complex social phenomena~\citep{salganik2020measuring}. Moreover, several orthogonal factors influence the outcome of the debate including the author's reputation, the quality of argumentation, the discussion topic, and the societal and historical context around the discussion. As studies accumulate evidence supporting the role of different factors in opinion change~\cite{tan2016winning}, future work can incorporate the social dimensions into more comprehensive and predictive models.

Second, the classifiers we use to detect political posts and to extract the social dimensions are accurate, yet not exhaustive nor error-free. The social dimension classifier is based on a conceptualization of social relationships that is broader than existing theoretical models~\citep{wellman1990different,fiske1992four,spencer2018rethinking}, and its proponents have shown empirically~\citep{choi2020ten} that it accounts for key dimensions of traditional psycholinguistic models~\citep{plutchik1980general,tausczik2010psychological}. Yet, future research should strive for models that are more accurate (i.e., lower error rate), more comprehensive (i.e., more dimensions), and especially more detailed (i.e., different nuances of a given dimension). In fact, many of the social dimensions that we use in our work have been charcterized by previous research as complex superpositions of different psychological constructs; for example, trust comprises both cognitive and affective components~\cite{dowell2015changing} that the tool we use in this work is not able to disentangle. Given the rapid progress of natural language processing technologies, we expect that improved models for language understanding can soon replace the ones used in this study. New methods could also attempt to qualify social dimensions in ways that we have not considered in this study, for example accounting for the \emph{directionality} of the social intent: our classifier detects the presence of a social intent in an utterance but cannot determine who is the subject that expresses the intent (e.g., \emph{``you trust me''} and \emph{``I trust you''} are both classified as trust, and are considered equivalent in our analysis).

Third, \rcmv is a platform with unique qualities: it is designed to attract members who are keen on participating in public discussions and who are open to change their opinion; it also provides clear way to track interactions and opinion changes. It is therefore likely that participants self-select to be particularly open to mutual understanding. On the one hand, these properties---akin to an idealized setting for communication---allow us to work on clean data, while preserving a good degree of ecological validity compared to studies conducted in the lab. On the other hand, replicating our experimental setup on multiple platforms is needed to support the generality of our findings.

Fourth, we analyze conversations with one exchange only---one post and one response---mainly because these make for the vast majority of interactions on \rcmv. With more extensive data at hand, future studies could look into how the role of different social dimensions changes as the dialogue progresses. More broadly, our results can provide a basis for operationalizing complex psychological theories of communication (e.g., transactional analysis~\cite{stewart1987ta}).

Fifth, and last, one of the main goals of this work is to propose a framework that can overcome the oversimplified design of opinion dynamics models by adding nuance to the types of social interactions considered. In most opinion dynamics models, interactions are either binarized, or associated with a polarity, to represent interactions with positive/negative sentiment, or between actors with same/different ideological stances~\cite{grabisch2020survey}. Our regression results show that neither sentiment nor political side explain opinion change as well as speaker intent, as operationalized by the social dimensions exchanged.

\smallskip

\clearpage

\section{Materials and methods}

\begin{table*}[t]
    \centering
    \small
    \caption{ %
    Explaination of the confounders used in the logistic regression models (Table~\ref{tab:homophily-log-reg}). More details are provided in Section~\ref{sec:confounders}.}
    \label{tab:homophily-log-reg-explaination}

\begin{tabular}{l|l|p{30em}}
\toprule
\multicolumn{2}{c}{\textbf{Variable}} & \multicolumn{1}{c}{\textbf{Description}} \\ \midrule
Sentiment   & Sentiment Pos.                  \dotfill & \multirow{2}{30em}{%
Positive and negative sentiment in $[0, 1]$ of the \rcmv comment, as extracted by Vader~\cite{hutto2014vader}.} \\
            & Sentiment Neg.                  \dotfill &          \\ \addlinespace
\multirow{1}{6em}{Political side (individual)} & Comment Left-Wing         \dotfill & \multirow{4}{30em}{%
Boolean variables describing whether the author of the comment (resp. post) ever participated in a subreddit that we identified as left-wing (resp. right-wing) at the time of their submission to \rcmv. } \\
            & Comment Right-Wing      \dotfill &          \\
            & Post Left-Wing                  \dotfill &          \\
            & Post Right-Wing                 \dotfill &          \\ \addlinespace
\multirow{1}{6em}{Political side (interaction)} & Both Polarized          \dotfill & %
Equal to $1$ if both the author of the comment and the author of the post ever participated in one of the subreddits that we identified as left-wing or right-wing. \\ % at the time of their submission to \rcmv. \\
            & Both Polarized \& Diff. Side \dotfill & % 
Equal to $1$ if the author of the comment participated in a left-wing subreddit and the author of the post in a right-wing subreddit, or vice-versa. \\
            & Diff. Side                      \dotfill & %
Equal to $1$ if the author of the comment participated in a left-wing (or right-wing) subreddit and the author of the post did not, or vice-versa. \\ \addlinespace
Political group   & Shared Group              \dotfill & %
Equal to $1$ if the author of the comment and that of the post ever participated in the same polarized subreddit. \\
\bottomrule
\end{tabular}
\end{table*}

\subsection{Classification of sociopolitical posts}
\label{sec:sociopolitical-classifier}

To focus on a homogeneous set of posts, we develop a supervised classifier that recognizes posts with a sociopolitical topic, according to the definition given by Moy and Gastil~\citep{moy2006predicting}. In addition, given the focus on opinion change, we wish to exclude posts that discuss strictly factual statements. To train such a classifier, we manually categorize the 2000 most popular (non-access-restricted) subreddits in 2019 as sociopolitical or not by looking at their description and a sample of their posts, and find $51$ sociopolitical subreddits. Then, we use such classification to build a training set for our supervised classifier. Specifically, we take a random sample of 50 posts per month for each sociopolitical subreddit, from 2011 to the end of 2019; if a subreddit does not have at least 50 posts in a month, we take all available posts for that month. We also take a random sample of equal total size for each month from all the non-sociopolitical subreddits. This way, we obtain a training set composed by \num{104292} sociopolitical posts (stratified over subreddits and time) and the same number of non-sociopolitical ones (stratified over time). Using the same sampling procedure, we collect a test set of equal size. We use the training set to train a logistic regression model with L1 regularization, to distinguish between sociopolitical and non-sociopolitical posts. As features, we employ $\{1, 2, 3\}$-grams, and keep only the \num{10000} most frequent ones in the whole dataset.

We evaluate the results of the classifier in two ways. First, on the test set, on which the classifier gets an average F1 score of $89.5\%.$ (detailed results in Table~SI1 and Table~SI2).

Then, we manually build a validation set of \rcmv posts by randomly selecting 500 posts from the subreddit and labelling each one as sociopolitical or not by manual inspection, according to the definition given before (Table~SI3 shows an excerpt). Out of the $500$ posts, $269$ are labelled as sociopolitical ($53.8\%$). On this validation set, our classifier obtains an F1-score of $75\%$ if we consider all posts, including the ones for which the main text of the post was subsequently removed by the author and only the title is available. If we consider only the $206$ posts where text is present ($120$ of which sociopolitical), the classifier obtains an F1-score of $82\%$. We report other accuracy metrics in Table~SI2. For this reason, in the rest of this work we consider only posts with text present. We then proceed to classify all \rcmv posts by using our sociopolitical classifier. Finally, we extract the comments of the posts that are recognized as sociopolitical.

To check how the precision of this classifier impacts our results, we repeat all our experiments using a different threshold to classify sociopolitical posts. Specifically, in this alternative setting we choose to classify posts as sociopolitical if the classifier assigns them a score of $0.75$ (instead of $0.5$). This experiment leads to \num{40296} posts classified as sociopolitical instead of \num{46046} (i.e., $87.5\%$). Also, by using this higher threshold on the validation data set, we obtain a higher precision of $83\%$ ($+7\%$) and a lower recall of $71\%$ ($-17\%$). Then, we repeat all of the experiments presented in this work. We observe that none of our results change substantially: for example, in Table~\ref{tab:homophily-log-reg} the significance of results stays unchanged, and the odds ratios change only in the third decimal digit in most cases (see Table SI8).

\subsection{Extracting social dimensions from text}
\label{sec:methods:10dims}

To extract the social dimensions from the set of sociopolitical posts and their respective comments, we leverage a previously developed model~\cite{choi2020ten} with a publicly-available Python implementation (\url{http://www.github.com/lajello/tendimensions})

Given a textual message $m$ and a social dimension $d$, the model estimates the likelihood that $m$ conveys $d$ by giving a score from 0 (least likely) to 1 (most likely). The model is not a multi-class classifier, rather it includes a set of independently-trained binary classifiers $C_d$, one per each dimension, i.e., it is a multi-label classifier. This choice is driven by the theoretical interpretation of the social dimensions~\cite{deri2018coloring}, as any sentence may potentially convey several dimensions at once (e.g., a message expressing both trust and emotional support). Each classifier is implemented by using a Long Short-Term Memory neural network (LSTM)~\cite{hochreiter1997long}, a type of Recurrent Neural Network (RNN) that is particularly effective in modeling both long and short-range semantic dependencies between words in a text, and it is therefore widely used in a variety of natural language processing tasks~\cite{sundermeyer2012lstm}. Similarly to most RNNs, LSTM accepts fixed-size inputs. This particular model takes in input a $300$-dimensional embedding vector of a word, one word at a time for all the words in the input text. Embedding vectors are dense numerical representations of the position of a word in a multidimensional semantic space learned from large text corpora. This model uses GloVe embeddings~\cite{pennington2014glove} learned from Common Crawl, a text corpus which contains $840$B tokens.

The dimensions classifiers $C_d$ are trained by using about 9k sentences manually labeled by trained crowdsourcing workers. Most of these sentences are taken from Reddit, which makes it the ideal platform to apply the model on. The models achieve very good classification performance which averages to an Area Under the Curve (AUC) of $0.84$ across dimensions. AUC is a standard performance metric that assesses the ability of a classifier to rank positive and negative instances by their likelihood score, independent of any fixed decision threshold (the AUC of a random classifier is expected to be 0.5, whereas the maximum value is 1).

In practice, the classifier estimates a score for each sentence $S$ in $m$ and returns the maximum score, namely: $s_d(m) = \max_{S \in m} s_d(S)$. By using the maximum score, we consider a message as likely to express dimension $d$ as its most likely sentence, thus avoiding the dilution effect of the average. This reflects the theoretical interpretation of the use of the social dimensions in language~\cite{deri2018coloring}: a dimension is conveyed effectively through language even when expressed only briefly.

\subsection{Binarization and normalization of social dimension scores}
\label{sec:normalization-scores}
To conduct our analysis, we binarize the classifier scores $s_d(m)$ via an indicator function that assigns dimension $d$ to $m$ if $s_d(m)$ is above a certain threshold $\theta_d$: 

\begin{equation}
    d(m) = 
    \begin{cases}
      1, & \text{if}\ s_d(m) \geq \theta_d\\
      0, & \text{otherwise}
    \end{cases}
 \end{equation}

We use dimension-specific thresholds because the empirical distribution of the classifier scores $s_d$ varies across dimensions, which makes the use of a fixed common threshold unpractical. We conservatively pick the value of $\theta_d$ as the $85^{th}$ percentile of the empirical distribution of the scores $s_d$, thus favoring high precision over recall. This effectively reduces the number of messages marked with each dimension to 15\% of the total number of messages.

When assigning a dimension $d$ to a message based on the single sentence that most prominently expresses that dimension, the probability of being labeled with $d$ naturally increases with the length of the message. Figure~SI3 shows that such increase is roughly linear with the number of words. To mitigate the length bias, we design a length-discounting factor. The typical length of messages may vary considerably across dimensions (Figure~SI2), therefore, to avoid excessively penalizing some dimensions over others, we use a dimension-specific discounting factor. Given a message $m$ with length $len(m)$ (measured in number of words), and labeled with dimension $d$ (i.e., such that $d(m) = 1$), we proceed as follows. First, we standardize $len(m)$ with respect to the length distribution of all messages labeled with $d$: given $\mu_{len}(d)$ and $\sigma_{len}(d)$ as the average and standard deviation of the length distribution of messages with $d$, the standardized value of length is $zlen_d(m) = \frac{len(m) - \mu_{len}(d)}{\sigma_{len}(d)}$.
We then redefine $d(m)$ as follows:

\begin{equation}
    d(m) = 
    \begin{cases}
      \frac{1}{1+ zlen_d(m)} & \text{if} \ s_d(m) \geq \theta_d \wedge zlen_d(m) >= 0 \\
      2 - \frac{1}{1-zlen_d(m)} & \text{if} \ s_d(m) \geq \theta_d \wedge zlen_d(m) < 0 \\
      0, & \text{if} \ s_d(m) < \theta_d
    \end{cases}
 \end{equation}

This length-discounted value is bounded between 0 and 2.
It is equal to 1 when the message length is equal to the average length of messages with dimension $d$.
It approaches 0 as the length of the message increases, and it gets closer to 2 as the length approaches 0, thus effectively weighting more those messages whose length is shorter than what is expected for the dimension considered (and vice versa). In Figure~SI4, we show the effect that the weight discounting has in reducing the cross-correlation between pairs of dimensions (i.e., in increasing their orthogonality).

\subsection{Odds ratios}
\label{sec:odds-ratios-definitions}

The length-discounted prior probability of a message (either a post or a comment) being labeled with dimensions $d$ is 

\begin{equation}
p(d) = \frac{\sum_{m \in {M}}{d(m)}}{2 \cdot |M|},
\end{equation}

where $M$ is the set of messages and the factor 2 is used to limit $p(d)$ between 0 and 1, as the values of $d(m)$ range from 0 to 2.

We define the conditional probability that a comment contains $d$, given that it received a $\Delta$ as

\begin{equation}
p(d | \Delta) = \frac{\sum_{m \in C_{\Delta}}{d(m)}}{2 \cdot |C_{\Delta}|},
\end{equation}

where $C_{\Delta}$ is the set of comments with $\Delta$. We use an analogous formulation for the set of messages $\noDelta$. The odds ratio between $d$ and $\Delta$, which is a measure of the strength of their association, is defined as

\begin{equation}
OR(p(d | \Delta), p(d | \noDelta)) = \frac { p(d | \Delta) / (1 - p(d | \Delta)) } { p(d | \noDelta) / (1 - p(d | \noDelta))}.
\end{equation}

We compute the conditional probability of a comment containing dimension $d_i$, given that its corresponding post contains dimension $d_j$, as

\begin{equation}
p(d_i(comment) | d_j(post)) = \frac{ \sum_{c \in C(P_{d_j}) }{d_i(c)}  }{  2 \cdot |C(P_{d_j})| },
\end{equation}

where $P_d$ is the set of posts with $d$, $C(P_d)$ is the corresponding set of comments. When considering only the set $C_{\Delta}$ of comments with $\Delta$ (and equivalently for $\noDelta$), the formula becomes

\begin{equation}
p_{\Delta}(d_i | d_j) = \frac{ \sum_{c \in C_{\Delta}(P_{d_j}) }{d_i(c)}  }{  2 \cdot |C_{\Delta}(P_{d_j})| }.
\end{equation}

As the joint distribution of the dimensions in comments and posts is different between the case of $\Delta$ and $\noDelta$,  to make the $p_{\Delta}(d_i | d_j)$ and $p_{\noDelta}(d_i | d_j)$ values comparable, we offset them by their expected value under a randomized null model. Specifically, we create a shuffled dataset via a random permutation $r$, so that the association between posts and comments is randomized, but the dimension-message association is unchanged. This null model destroys the association between posts and comments, thus reflecting the baseline probability that a post with $d_j$ would receive a comment with $d_i$ just by chance. We calculate $p^{r}_{\Delta}(d_i | d_j)$ and $p^{r}_{\noDelta}(d_i | d_j)$ in this randomized model, and we then calculate the odd ratios between the real data and the random baseline:

\begin{equation}
OR(p_{\Delta}(d_i | d_j),  p^{r}_{\Delta}(d_i | d_j) ) = \frac { p_{\Delta}(d_i | d_j) / (1 - p_{\Delta}(d_i | d_j)) } { p^{r}_{\Delta}(d_i | d_j) / (1 - p^{r}_{\Delta}(d_i | d_j))}.
\end{equation}

We indicate this odds ratio with $OR_{\Delta}(d_i, d_j)$, and we analogously define $OR_{\noDelta}(d_i, d_j)$.

The $95\%$ confidence intervals of the odds ratios are calculated as:
\begin{equation}
ci = 1.96 \cdot \sqrt{ \frac{1}{|C_{d,\Delta}|} + \frac{1}{|C_{d,\noDelta}|} + \frac{1}{|C_{\overbar{d},\Delta}|} + \frac{1}{|C_{\overbar{d},\noDelta}|}},
\end{equation}
where 1.96 is the critical value of the Normal distribution at $\alpha/2$ (with $\alpha = 0.05$) and $|C_{\bullet,\bullet}|$ represents the cardinality of the set of comments with or without a given dimension ($d$ or $\overbar{d}$) and with or without a delta ($\Delta$ or $\noDelta$).

\subsection{Confounders}
\label{sec:confounders}

Beside the social dimensions, we test possible confounders in our regression models. To model message-level confounders, we include in our models the positive and negative sentiment of the message as measured by Vader~\cite{hutto2014vader}. We also consider the length of the message $len(m)$ (Table~); specifically, we use the standardized value of $\log(len(m))$, to account for the fat-tailed distribution of message lengths.

To estimate the ideological leaning of the participants to the conversation, we consider subreddits in which a user has participated with a submission at any point in time \emph{before} their message in \rcmv under consideration. From the posting history of each pair of poster and commenter, we extract three sets of variables that capture, respectively, the political leaning of each of the two users, the interaction between their leanings, and their similarity.

The first set describes whether each of the two users has participated in a right-wing or a left-wing subreddit. We manually select a set of $10$ right-leaning subreddits (e.g., \R{The\_Donald}, \R{Conservative}) and a set of $15$ left-leaning subreddits (e.g., \R{SandersForPresident}, \R{Socialism}). In the selected groups, rules typically suggest that participants adhere to the ideological view of the subreddits; as such, participation can be taken as a meaningful proxy of a political ideology.

The second set expresses whether both involved users have participated in any politically-identified subreddit, and whether they participated on the same side. We use two boolean variables and their interaction to encode this information.

The last set considers whether the two users have participated in precisely the same subreddit among the $25$ subreddits selected to assess political leaning, plus a list of $14$ additional subreddits whose members express homogeneous political views, but that are not necessarily well-positioned on the traditional left-right spectrum (e.g., \R{atheism}, \R{conspiracy}, or \R{NeutralPolitics}).

Table~\ref{tab:homophily-log-reg-explaination} summarizes all the considered confounders.

\section*{Acknowledgements}
{LMA acknowledges the support from the Carlsberg Foundation through the COCOONS project (CF21-0432). The funder had no role in study design, data collection and analysis, decision to publish, or preparation of the manuscript.}

\section*{Data availability}
The raw Reddit data is freely available through the \url{pushift.io} API. The data that we used to for the experiments, and the sociopolitical classifier are available at: \url{https://github.com/corradomonti/10-dim-of-op-change}. The social dimensions classifier is available at: \url{https://github.com/lajello/tendimensions}. 

\section*{Author contributions}
CM collected the data, CM and LMA conducted the experiments. CM, LMA, GDFM, and FB conceived the experiments, wrote and reviewed the manuscript.

\section*{Competing interests}

The authors declare no competing interests.

%\bibliography{references}

\begin{thebibliography}{10}
\urlstyle{rm}
\expandafter\ifx\csname url\endcsname\relax
  \def\url#1{\texttt{#1}}\fi
\expandafter\ifx\csname urlprefix\endcsname\relax\def\urlprefix{URL }\fi
\expandafter\ifx\csname doiprefix\endcsname\relax\def\doiprefix{DOI: }\fi
\providecommand{\bibinfo}[2]{#2}
\providecommand{\eprint}[2][]{\url{#2}}

\bibitem{habermas1962structural}
\bibinfo{author}{Habermas, J.}
\newblock \emph{\bibinfo{title}{The {{Structural Transformation}} of the
  {{Public Sphere}}: An {{Inquiry}} into a {{Category}} of {{Bourgeois
  Society}}}} (\bibinfo{publisher}{{MIT Press}}, \bibinfo{address}{{Cambridge,
  Mass}}, \bibinfo{year}{1962}).

\bibitem{habermas1992further}
\bibinfo{author}{Habermas, J.}
\newblock \bibinfo{journal}{\bibinfo{title}{Further reflections on the public
  sphere}}.
\newblock {\emph{\JournalTitle{Habermas and the public sphere}}}
  \textbf{\bibinfo{volume}{428}} (\bibinfo{year}{1992}).

\bibitem{gimmler2001deliberative}
\bibinfo{author}{Gimmler, A.}
\newblock \bibinfo{journal}{\bibinfo{title}{Deliberative democracy, the public
  sphere and the internet}}.
\newblock {\emph{\JournalTitle{Philosophy \& Social Criticism}}}
  \textbf{\bibinfo{volume}{27}}, \bibinfo{pages}{21--39}
  (\bibinfo{year}{2001}).

\bibitem{fuchs2014social}
\bibinfo{author}{Fuchs, C.}
\newblock \bibinfo{journal}{\bibinfo{title}{Social media and the public
  sphere}}.
\newblock {\emph{\JournalTitle{tripleC: Communication, Capitalism \& Critique.
  Open Access Journal for a Global Sustainable Information Society}}}
  \textbf{\bibinfo{volume}{12}}, \bibinfo{pages}{57--101}
  (\bibinfo{year}{2014}).

\bibitem{habermas1981lifeworld}
\bibinfo{author}{Habermas, J.}
\newblock \emph{\bibinfo{title}{Lifeworld and {{System}}: A {{Critique}} of
  {{Functionalist Reason}}}}.
\newblock No. \bibinfo{number}{J\"urgen Habermas. Transl. by Thomas MacCarthy ;
  Vol. 2} in \bibinfo{series}{The {{Theory}} of {{Communicative Action}}}
  (\bibinfo{publisher}{{Beacon}}, \bibinfo{address}{{Boston}},
  \bibinfo{year}{1981}).

\bibitem{habermas1979communication}
\bibinfo{author}{Habermas, J.}
\newblock \emph{\bibinfo{title}{Communication and the {{Evolution}} of
  {{Society}}}}, vol. \bibinfo{volume}{572} (\bibinfo{publisher}{{Beacon
  Press}}, \bibinfo{year}{1979}).

\bibitem{habermas1981reason}
\bibinfo{author}{Habermas, J.}
\newblock \emph{\bibinfo{title}{Reason and the {{Rationalization}} of
  {{Society}}}}.
\newblock No. \bibinfo{number}{J\"urgen Habermas. Transl. by Thomas MacCarthy ;
  Vol. 1} in \bibinfo{series}{The {{Theory}} of {{Communicative Action}}}
  (\bibinfo{publisher}{{Beacon}}, \bibinfo{address}{{Boston}},
  \bibinfo{year}{1981}).

\bibitem{krauss1998language}
\bibinfo{author}{Krauss, R.~M.} \& \bibinfo{author}{Chiu, C.-Y.}
\newblock \bibinfo{title}{Language and {{Social Behavior}}}.
\newblock In \emph{\bibinfo{booktitle}{Handbook of {{Social Psychology}}}},
  vol.~\bibinfo{volume}{2} (\bibinfo{publisher}{{McGraw-Hill}},
  \bibinfo{year}{1998}), \bibinfo{edition}{fourth} edn.

\bibitem{austin1962how}
\bibinfo{author}{Austin, J.~L.}
\newblock \emph{\bibinfo{title}{How to {{Do Things}} with {{Words}}: The
  {{William James Lectures}} Delivered at {{Harvard University}} in 1955}}
  (\bibinfo{publisher}{{Harvard Univ. Press}}, \bibinfo{address}{{Cambridge,
  Mass}}, \bibinfo{year}{1962}).

\bibitem{grabisch2020survey}
\bibinfo{author}{Grabisch, M.} \& \bibinfo{author}{Rusinowska, A.}
\newblock \bibinfo{journal}{\bibinfo{title}{A survey on nonstrategic models of
  opinion dynamics}}.
\newblock {\emph{\JournalTitle{Games}}} \textbf{\bibinfo{volume}{11}},
  \bibinfo{pages}{65} (\bibinfo{year}{2020}).

\bibitem{choi2020ten}
\bibinfo{author}{Choi, M.}, \bibinfo{author}{Aiello, L.~M.},
  \bibinfo{author}{Varga, K.~Z.} \& \bibinfo{author}{Quercia, D.}
\newblock \bibinfo{title}{Ten social dimensions of conversations and
  relationships}.
\newblock In \emph{\bibinfo{booktitle}{Proceedings of the Web Conference
  2020}}, \bibinfo{pages}{1514--1525} (\bibinfo{year}{2020}).

\bibitem{Rosenberg2005}
\bibinfo{author}{Rosenberg, S.}
\newblock \bibinfo{journal}{\bibinfo{title}{The empirical study of deliberative
  democracy: Setting a research agenda}}.
\newblock {\emph{\JournalTitle{Acta Politica}}} \textbf{\bibinfo{volume}{40}},
  \bibinfo{pages}{212--224}, \doiprefix\url{10.1057/palgrave.ap.5500105}
  (\bibinfo{year}{2005}).

\bibitem{habermas2018interview}
\bibinfo{author}{Habermas, J.}
\newblock \bibinfo{title}{Interview with {J{\"u}rgen} {Habermas}}.
\newblock In \bibinfo{editor}{B{\"a}chtiger, A.}, \bibinfo{editor}{Dryzek,
  J.~S.}, \bibinfo{editor}{Mansbridge, J.} \& \bibinfo{editor}{Warren, M.}
  (eds.) \emph{\bibinfo{booktitle}{The {Oxford} {Handbook} of {Deliberative}
  {Democracy}}}, \bibinfo{pages}{870--882},
  \doiprefix\url{10.1093/oxfordhb/9780198747369.013.60}
  (\bibinfo{publisher}{Oxford University Press}, \bibinfo{year}{2018}).

\bibitem{heng2003habermas}
\bibinfo{author}{Heng, M.~S.} \& \bibinfo{author}{De~Moor, A.}
\newblock \bibinfo{journal}{\bibinfo{title}{From habermas's communicative
  theory to practice on the internet}}.
\newblock {\emph{\JournalTitle{Information Systems Journal}}}
  \textbf{\bibinfo{volume}{13}}, \bibinfo{pages}{331--352}
  (\bibinfo{year}{2003}).

\bibitem{blau64exchange}
\bibinfo{author}{Blau, P.~M.}
\newblock \emph{\bibinfo{title}{Exchange and Power in Social Life}}
  (\bibinfo{publisher}{{Transaction Publishers}}, \bibinfo{year}{1964}).

\bibitem{moy2006predicting}
\bibinfo{author}{Moy, P.} \& \bibinfo{author}{Gastil, J.}
\newblock \bibinfo{journal}{\bibinfo{title}{Predicting deliberative
  conversation: The impact of discussion networks, media use, and political
  cognitions}}.
\newblock {\emph{\JournalTitle{Political Communication}}}
  \textbf{\bibinfo{volume}{23}}, \bibinfo{pages}{443--460}
  (\bibinfo{year}{2006}).

\bibitem{deri2018coloring}
\bibinfo{author}{Deri, S.}, \bibinfo{author}{Rappaz, J.},
  \bibinfo{author}{Aiello, L.~M.} \& \bibinfo{author}{Quercia, D.}
\newblock \bibinfo{journal}{\bibinfo{title}{Coloring in the {{Links}}:
  Capturing {{Social Ties}} as {{They}} are {{Perceived}}}}.
\newblock {\emph{\JournalTitle{Proceedings of the ACM on Human-Computer
  Interaction}}} \textbf{\bibinfo{volume}{2}}, \bibinfo{pages}{1--18},
  \doiprefix\url{10.1145/3274312} (\bibinfo{year}{2018}).
\newblock \eprint{1902.04528}.

\bibitem{wellman1990different}
\bibinfo{author}{Wellman, B.} \& \bibinfo{author}{Wortley, S.}
\newblock \bibinfo{journal}{\bibinfo{title}{Different strokes from different
  folks: Community ties and social support}}.
\newblock {\emph{\JournalTitle{American journal of Sociology}}}
  \textbf{\bibinfo{volume}{96}}, \bibinfo{pages}{558--588}
  (\bibinfo{year}{1990}).

\bibitem{fiske1992four}
\bibinfo{author}{Fiske, A.~P.}
\newblock \bibinfo{journal}{\bibinfo{title}{The four elementary forms of
  sociality: Framework for a unified theory of social relations.}}
\newblock {\emph{\JournalTitle{Psychological review}}}
  \textbf{\bibinfo{volume}{99}}, \bibinfo{pages}{689} (\bibinfo{year}{1992}).

\bibitem{spencer2018rethinking}
\bibinfo{author}{Spencer, L.} \& \bibinfo{author}{Pahl, R.}
\newblock \emph{\bibinfo{title}{Rethinking Friendship}}
  (\bibinfo{publisher}{{Princeton University Press}}, \bibinfo{year}{2018}).

\bibitem{feder2021causal}
\bibinfo{author}{Feder, A.} \emph{et~al.}
\newblock \bibinfo{journal}{\bibinfo{title}{Causal inference in natural
  language processing: Estimation, prediction, interpretation and beyond}}.
\newblock {\emph{\JournalTitle{arXiv preprint arXiv:2109.00725}}}
  (\bibinfo{year}{2021}).

\bibitem{fiske2007universal}
\bibinfo{author}{Fiske, S.~T.}, \bibinfo{author}{Cuddy, A.~J.} \&
  \bibinfo{author}{Glick, P.}
\newblock \bibinfo{journal}{\bibinfo{title}{Universal dimensions of social
  cognition: Warmth and competence}}.
\newblock {\emph{\JournalTitle{Trends in cognitive sciences}}}
  \textbf{\bibinfo{volume}{11}}, \bibinfo{pages}{77--83}
  (\bibinfo{year}{2007}).

\bibitem{luhmann1982trust}
\bibinfo{author}{Luhmann, N.}
\newblock \emph{\bibinfo{title}{Trust and Power}} (\bibinfo{publisher}{{John
  Wiley \& Sons}}, \bibinfo{year}{1982}).

\bibitem{mcpherson2001birds}
\bibinfo{author}{McPherson, M.}, \bibinfo{author}{{Smith-Lovin}, L.} \&
  \bibinfo{author}{Cook, J.~M.}
\newblock \bibinfo{journal}{\bibinfo{title}{Birds of a feather: Homophily in
  social networks}}.
\newblock {\emph{\JournalTitle{Annual review of sociology}}}
  \textbf{\bibinfo{volume}{27}}, \bibinfo{pages}{415--444}
  (\bibinfo{year}{2001}).

\bibitem{tajfel2010social}
\bibinfo{author}{Tajfel, H.}
\newblock \emph{\bibinfo{title}{Social Identity and Intergroup Relations}}
  (\bibinfo{publisher}{{Cambridge University Press}}, \bibinfo{year}{2010}).

\bibitem{argyle2013psychology}
\bibinfo{author}{Argyle, M.}
\newblock \emph{\bibinfo{title}{The Psychology of Happiness}}
  (\bibinfo{publisher}{{Routledge}}, \bibinfo{year}{2013}).

\bibitem{tajfel1979integrative}
\bibinfo{author}{Tajfel, H.}, \bibinfo{author}{Turner, J.~C.},
  \bibinfo{author}{Austin, W.~G.} \& \bibinfo{author}{Worchel, S.}
\newblock \bibinfo{journal}{\bibinfo{title}{An integrative theory of intergroup
  conflict}}.
\newblock {\emph{\JournalTitle{Organizational Identity}}}
  (\bibinfo{year}{1979}).

\bibitem{massachs2020roots}
\bibinfo{author}{Massachs, J.}, \bibinfo{author}{Monti, C.},
  \bibinfo{author}{De~Francisci~Morales, G.} \& \bibinfo{author}{Bonchi, F.}
\newblock \bibinfo{title}{Roots of {{Trumpism}}: Homophily and {{Social
  Feedback}} in {{Donald Trump Support}} on {{Reddit}}}.
\newblock In \emph{\bibinfo{booktitle}{{{WebSci}} '20: 12th International
  {{ACM}} Web Science Conference}} (\bibinfo{year}{2020}).

\bibitem{benjamini1995controlling}
\bibinfo{author}{Benjamini, Y.} \& \bibinfo{author}{Hochberg, Y.}
\newblock \bibinfo{journal}{\bibinfo{title}{Controlling the false discovery
  rate: A practical and powerful approach to multiple testing}}.
\newblock {\emph{\JournalTitle{Journal of the Royal statistical society: series
  B (Methodological)}}} \textbf{\bibinfo{volume}{57}},
  \bibinfo{pages}{289--300} (\bibinfo{year}{1995}).

\bibitem{keltner1997defending}
\bibinfo{author}{Keltner, D.} \& \bibinfo{author}{Robinson, R.~J.}
\newblock \bibinfo{journal}{\bibinfo{title}{Defending the status quo: Power and
  bias in social conflict}}.
\newblock {\emph{\JournalTitle{Personality and Social Psychology Bulletin}}}
  \textbf{\bibinfo{volume}{23}}, \bibinfo{pages}{1066--1077}
  (\bibinfo{year}{1997}).

\bibitem{gouldner1960norm}
\bibinfo{author}{Gouldner, A.~W.}
\newblock \bibinfo{journal}{\bibinfo{title}{The norm of reciprocity: A
  preliminary statement}}.
\newblock {\emph{\JournalTitle{American sociological review}}}
  \bibinfo{pages}{161--178} (\bibinfo{year}{1960}).

\bibitem{terry1997habermas}
\bibinfo{author}{Terry, P.~R.}
\newblock \bibinfo{journal}{\bibinfo{title}{Habermas and education: Knowledge,
  communication, discourse}}.
\newblock {\emph{\JournalTitle{Curriculum Studies}}}
  \textbf{\bibinfo{volume}{5}}, \bibinfo{pages}{269--279},
  \doiprefix\url{10.1080/14681369700200019} (\bibinfo{year}{1997}).

\bibitem{follett2011constructive}
\bibinfo{author}{Follett, M.~P.}
\newblock \bibinfo{journal}{\bibinfo{title}{Constructive conflict}}.
\newblock {\emph{\JournalTitle{Sociology of Organizations: Structures and
  Relationships}}} \textbf{\bibinfo{volume}{417}} (\bibinfo{year}{2011}).

\bibitem{cambria2017practical}
\bibinfo{author}{Cambria, E.}, \bibinfo{author}{Das, D.},
  \bibinfo{author}{Bandyopadhyay, S.} \& \bibinfo{author}{Feraco, A.}
\newblock \emph{\bibinfo{title}{A practical guide to sentiment analysis}}
  (\bibinfo{publisher}{Springer}, \bibinfo{year}{2017}).

\bibitem{bohman1999theories}
\bibinfo{author}{Bohman, J.}
\newblock \bibinfo{journal}{\bibinfo{title}{Theories, practices, and pluralism:
  A pragmatic interpretation of critical social science}}.
\newblock {\emph{\JournalTitle{Philosophy of the social sciences}}}
  \textbf{\bibinfo{volume}{29}}, \bibinfo{pages}{459--480}
  (\bibinfo{year}{1999}).

\bibitem{buyalskaya2021golden}
\bibinfo{author}{Buyalskaya, A.}, \bibinfo{author}{Gallo, M.} \&
  \bibinfo{author}{Camerer, C.~F.}
\newblock \bibinfo{journal}{\bibinfo{title}{The golden age of social science}}.
\newblock {\emph{\JournalTitle{Proceedings of the National Academy of
  Sciences}}} \textbf{\bibinfo{volume}{118}} (\bibinfo{year}{2021}).

\bibitem{steiner2005deliberative}
\bibinfo{author}{Steiner, J.}, \bibinfo{author}{B{\"a}chtiger, A.},
  \bibinfo{author}{Sp{\"o}rndli, M.} \& \bibinfo{author}{Steenbergen, M.~R.}
\newblock \emph{\bibinfo{title}{Deliberative politics in action. Analysing
  parliamentary discourse}} (\bibinfo{publisher}{{Cambridge University Press}},
  \bibinfo{year}{2005}).

\bibitem{steenbergen2003measuring}
\bibinfo{author}{Steenbergen, M.~R.}, \bibinfo{author}{B{\"a}chtiger, A.},
  \bibinfo{author}{Sp{\"o}rndli, M.} \& \bibinfo{author}{Steiner, J.}
\newblock \bibinfo{journal}{\bibinfo{title}{Measuring political deliberation: A
  discourse quality index}}.
\newblock {\emph{\JournalTitle{Comparative European Politics}}}
  \textbf{\bibinfo{volume}{1}}, \bibinfo{pages}{21--48} (\bibinfo{year}{2003}).

\bibitem{gerber2014deliberative}
\bibinfo{author}{Gerber, M.}, \bibinfo{author}{B{\"a}chtiger, A.},
  \bibinfo{author}{Fiket, I.}, \bibinfo{author}{Steenbergen, M.} \&
  \bibinfo{author}{Steiner, J.}
\newblock \bibinfo{journal}{\bibinfo{title}{Deliberative and non-deliberative
  persuasion: Mechanisms of opinion formation in europolis}}.
\newblock {\emph{\JournalTitle{European Union Politics}}}
  \textbf{\bibinfo{volume}{15}}, \bibinfo{pages}{410--429}
  (\bibinfo{year}{2014}).

\bibitem{salganik2020measuring}
\bibinfo{author}{Salganik, M.~J.} \emph{et~al.}
\newblock \bibinfo{journal}{\bibinfo{title}{Measuring the predictability of
  life outcomes with a scientific mass collaboration}}.
\newblock {\emph{\JournalTitle{Proceedings of the National Academy of
  Sciences}}} \textbf{\bibinfo{volume}{117}}, \bibinfo{pages}{8398--8403}
  (\bibinfo{year}{2020}).

\bibitem{tan2016winning}
\bibinfo{author}{Tan, C.}, \bibinfo{author}{Niculae, V.},
  \bibinfo{author}{{Danescu-Niculescu-Mizil}, C.} \& \bibinfo{author}{Lee, L.}
\newblock \bibinfo{journal}{\bibinfo{title}{Winning {{Arguments}}: Interaction
  {{Dynamics}} and {{Persuasion Strategies}} in {{Good}}-{{Faith Online
  Discussions}}}}.
\newblock {\emph{\JournalTitle{Proceedings of the 25th International Conference
  on World Wide Web - WWW '16}}} \bibinfo{pages}{613--624},
  \doiprefix\url{10.1145/2872427.2883081} (\bibinfo{year}{2016}).
\newblock \eprint{1602.01103}.

\bibitem{plutchik1980general}
\bibinfo{author}{Plutchik, R.}
\newblock \bibinfo{title}{A general psychoevolutionary theory of emotion}.
\newblock In \emph{\bibinfo{booktitle}{Theories of emotion}},
  \bibinfo{pages}{3--33} (\bibinfo{publisher}{Elsevier}, \bibinfo{year}{1980}).

\bibitem{tausczik2010psychological}
\bibinfo{author}{Tausczik, Y.~R.} \& \bibinfo{author}{Pennebaker, J.~W.}
\newblock \bibinfo{journal}{\bibinfo{title}{The psychological meaning of words:
  Liwc and computerized text analysis methods}}.
\newblock {\emph{\JournalTitle{Journal of language and social psychology}}}
  \textbf{\bibinfo{volume}{29}}, \bibinfo{pages}{24--54}
  (\bibinfo{year}{2010}).

\bibitem{dowell2015changing}
\bibinfo{author}{Dowell, D.}, \bibinfo{author}{Morrison, M.} \&
  \bibinfo{author}{Heffernan, T.}
\newblock \bibinfo{journal}{\bibinfo{title}{The changing importance of
  affective trust and cognitive trust across the relationship lifecycle: A
  study of business-to-business relationships}}.
\newblock {\emph{\JournalTitle{Industrial Marketing Management}}}
  \textbf{\bibinfo{volume}{44}}, \bibinfo{pages}{119--130}
  (\bibinfo{year}{2015}).

\bibitem{stewart1987ta}
\bibinfo{author}{Stewart, I.} \& \bibinfo{author}{Joines, V.}
\newblock \emph{\bibinfo{title}{{{TA}} Today: A New Introduction to
  Transactional Analysis}} (\bibinfo{publisher}{{Lifespace Pub.}},
  \bibinfo{year}{1987}).

\bibitem{hutto2014vader}
\bibinfo{author}{Hutto, C.~J.} \& \bibinfo{author}{Gilbert, E.}
\newblock \bibinfo{title}{Vader: A parsimonious rule-based model for sentiment
  analysis of social media text}.
\newblock In \emph{\bibinfo{booktitle}{Proceedings of the International
  {{AAAI}} Conference on Weblogs and Social Media}}, {{ICWSM}},
  \bibinfo{pages}{216--225} (\bibinfo{publisher}{{AAAI}},
  \bibinfo{year}{2014}).

\bibitem{hochreiter1997long}
\bibinfo{author}{Hochreiter, S.} \& \bibinfo{author}{Schmidhuber, J.}
\newblock \bibinfo{journal}{\bibinfo{title}{Long short-term memory}}.
\newblock {\emph{\JournalTitle{Neural computation}}}
  \textbf{\bibinfo{volume}{9}}, \bibinfo{pages}{1735--1780}
  (\bibinfo{year}{1997}).

\bibitem{sundermeyer2012lstm}
\bibinfo{author}{Sundermeyer, M.}, \bibinfo{author}{Schl{\"u}ter, R.} \&
  \bibinfo{author}{Ney, H.}
\newblock \bibinfo{title}{{{LSTM}} neural networks for language modeling}.
\newblock In \emph{\bibinfo{booktitle}{Thirteenth Annual Conference of the
  International Speech Communication Association}}, Interspeech
  (\bibinfo{year}{2012}).

\bibitem{pennington2014glove}
\bibinfo{author}{Pennington, J.}, \bibinfo{author}{Socher, R.} \&
  \bibinfo{author}{Manning, C.}
\newblock \bibinfo{title}{Glove: Global vectors for word representation}.
\newblock In \emph{\bibinfo{booktitle}{Proceedings of the Conference on
  Empirical Methods in Natural Language Processing}}, {{EMNLP}},
  \bibinfo{pages}{1532--1543} (\bibinfo{publisher}{{Association for
  Computational Linguistics}}, \bibinfo{year}{2014}).

\end{thebibliography}


\begin{thebibliography}{}
\urlstyle{rm}
\expandafter\ifx\csname url\endcsname\relax
  \def\url#1{\texttt{#1}}\fi
\expandafter\ifx\csname urlprefix\endcsname\relax\def\urlprefix{URL }\fi
\expandafter\ifx\csname doiprefix\endcsname\relax\def\doiprefix{DOI: }\fi
\providecommand{\bibinfo}[2]{#2}
\providecommand{\eprint}[2][]{\url{#2}}

\end{thebibliography}

\end{document}

\title{Supplementary Information}

\author[1]{Corrado Monti}
\author[2,3,*]{Luca Maria Aiello}
\author[1]{Gianmarco De Francisci Morales}
\author[1,4]{Francesco Bonchi}

\affil[1]{CENTAI, Torino, Italy}
\affil[2]{IT University of Copenhagen, Copenhagen, Denmark}
\affil[3]{Pioneer Centre for AI, Copenhagen, Denmark}
\affil[4]{Eurecat, Barcelona, Spain}

\flushbottom
\maketitle

\section*{Sociopolitical topic classification}

In order to extract our data set of posts with a sociopolitical topic from \rcmv, we build a supervised classifier. The development and training of such a classifier is described in Section~\ref{sec:sociopolitical-classifier} in \emph{Materials \& Methods}. Here, we report further information on the quality of this classifier.

First, we build a test set with the same procedure we used for the training data of the classifier; that is, we aggregate posts from different subreddits after we categorize each subreddit as sociopolitical or not (see Section~\ref{sec:sociopolitical-classifier} for details). On this test set, we obtain an average F1 score of $89.5\%$. However, since the activity on Reddit varies significantly in the nine years we consider, we further investigate whether the performance of this classifier changes over time. Table~\ref{table:sociopol-accuracy} reports the F1 score as measured on the posts from each year. We find that the quality in classification presents very limited variance over different years (i.e., within $\pm2.5$ p.p. from the global F1). Table~\ref{table:sociopol-metrics} reports additional metrics for this classifier.

Second, we build a validation set by selecting a random sample of \rcmv posts that we categorize as sociopolitical or not, according to the definition given by Moy and Gastil~\cite{moy2006predicting} (see \emph{Research Design} section, page~\pageref{sec:selecting-sociopol}). Table~\ref{table:sociopol-examples} reports a random excerpt of this validation data set, which provides concrete examples of the ``sociopolitical'' category. On this validation set, the classifier obtains an F1-score of $82\%$ when considering the posts where text is present in the body of the post. Table~\ref{table:sociopol-metrics} also reports additional metrics on this data set.

\begin{table}[b]
\caption{Classification performance of our sociopolitical classifier on the test set of Reddit posts over the years. In italic, years for which there are no posts in \rcmv. }
\label{table:sociopol-accuracy}
\centering
\footnotesize
\begin{tabular}{lllllllllll}
\toprule
\textbf{Year}     & \emph{2011} & \emph{2012} & 2013 & 2014 & 2015 & 2016 & 2017 & 2018 & 2019 \\
\cmidrule(lr){1-10}
\textbf{F1-score} & \emph{0.87} & \emph{0.89} & 0.89 & 0.89 & 0.90 & 0.92 & 0.91 & 0.90 & 0.89 \\
\bottomrule
\end{tabular}
\end{table}

\begin{table}[b]
\caption{Classification performance of the sociopolitical classifier on the automatically-built test set and on the human-evaluated validation set.}
\label{table:sociopol-metrics}
\centering
\footnotesize
\begin{tabular}{lllll}
\toprule
\textbf{Measure}  & \emph{F1-score} & \emph{Precision} & \emph{Recall} & \emph{Accuracy} \\
\midrule
\textbf{Test set } & \emph{0.90} & \emph{0.91} & \emph{0.88} & \emph{0.90} \\
\textbf{Validation set } & \emph{0.82} & \emph{0.76} & \emph{0.88} & \emph{0.77} \\
\bottomrule
\end{tabular}
\end{table}

\begin{table*}[]
  \caption{Examples from the validation set of ChangeMyView posts, hand-categorized as sociopolitical or not, according to our definition.}
  \label{table:sociopol-examples}
  \centering
  \footnotesize
\begin{tabular}{lp{14cm}}
\toprule
Sociopolitical & Title \\
\midrule
 Yes & CMV: The Second Amendment can no longer serve its intended purpose. \\
 Yes & I believe in the event of a major catastrophic natural disaster, the government should let its citizens enjoy blissful ignorance until the last moment. CMV. \\
 Yes & Fennisists and the metoo movement have caused more damage than good and are partially responsible for the rise of redpillers/mgtow \\
 No & CMV: In terms of overall health, one 12 oz Coca Cola per day is worse than one cigarette per day. \\
 Yes & CMV: Men should have a form of Abortion \\
 Yes & I do not condone abortion in any capacity. CMV \\
 Yes & I don't believe it is my responsibility to censor my actions or language to keep from offending others. CMV \\
 Yes & CMV: Theism and Atheism are both invalid arguments. \\
 No & CMV: Life is pointless \\
 No & CMV: Subreddits that disable downvoting should not be eligible to make the front page. \\
 Yes & CMV: Inciting Violence Should Be Protected Under Free Speech \\
 No & CMV: I believe OJ Simpson , by himself, committed the murders of Ron Goldman and Nicole Brown Simpson. \\
 Yes & CMV: America is not the greatest country in the world \\
 Yes & CMV: I don't think anyone should have to pay child support. Ever. \\
 Yes & CMV: Rewarding children for trying and participating (and winning) is the correct model; those who suggest just awarding the winners are wrong \\
 No & I believe the artificial fluoridation of water is safe. CMV. \\
 Yes & CMV: Many white people's social justice advocacy is self-serving and insincere. \\
 No & CMV: The Lottery is not real, and the winners are paid actors. \\
 Yes & CMV: Despite intuition to the contrary, I cannot see why a transracial person should not be accepted the way a transgender person is. \\
 Yes & CMV: One is able to criticize Islamic texts without being labelled racist \\
 No & CMV: /r/NoFap is bullshit \\
 No & I think Zombies are extremely overused and Dinosaurs need to make a comeback. CMV \\
\bottomrule
\end{tabular}

  \normalsize
\end{table*}

\section*{Data}

\begin{figure}
  \begin{center}
    \begin{subfigure}{0.5\linewidth}
      \centering
      \includegraphics[width=0.9\linewidth]{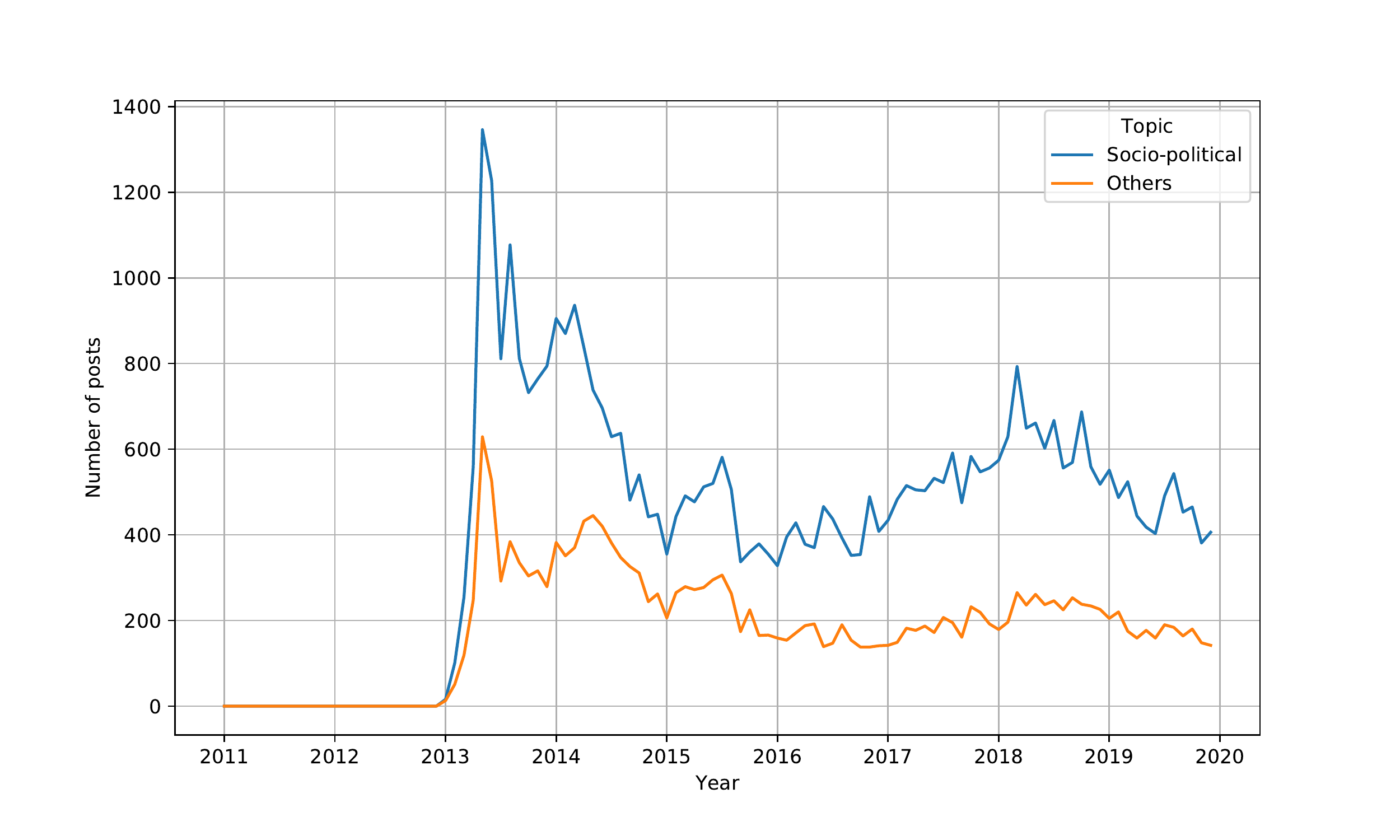}
      \caption{
      Number of sociopolitical and not-sociopolitical posts on \rcmv per month over time, as categorized by our supervised classifier.
      }
      \label{fig:sociopol-cmv-over-time}
    \end{subfigure}
  \end{center}
  \begin{subfigure}{0.45\linewidth}
    \centering
    \includegraphics[width=0.9\linewidth]{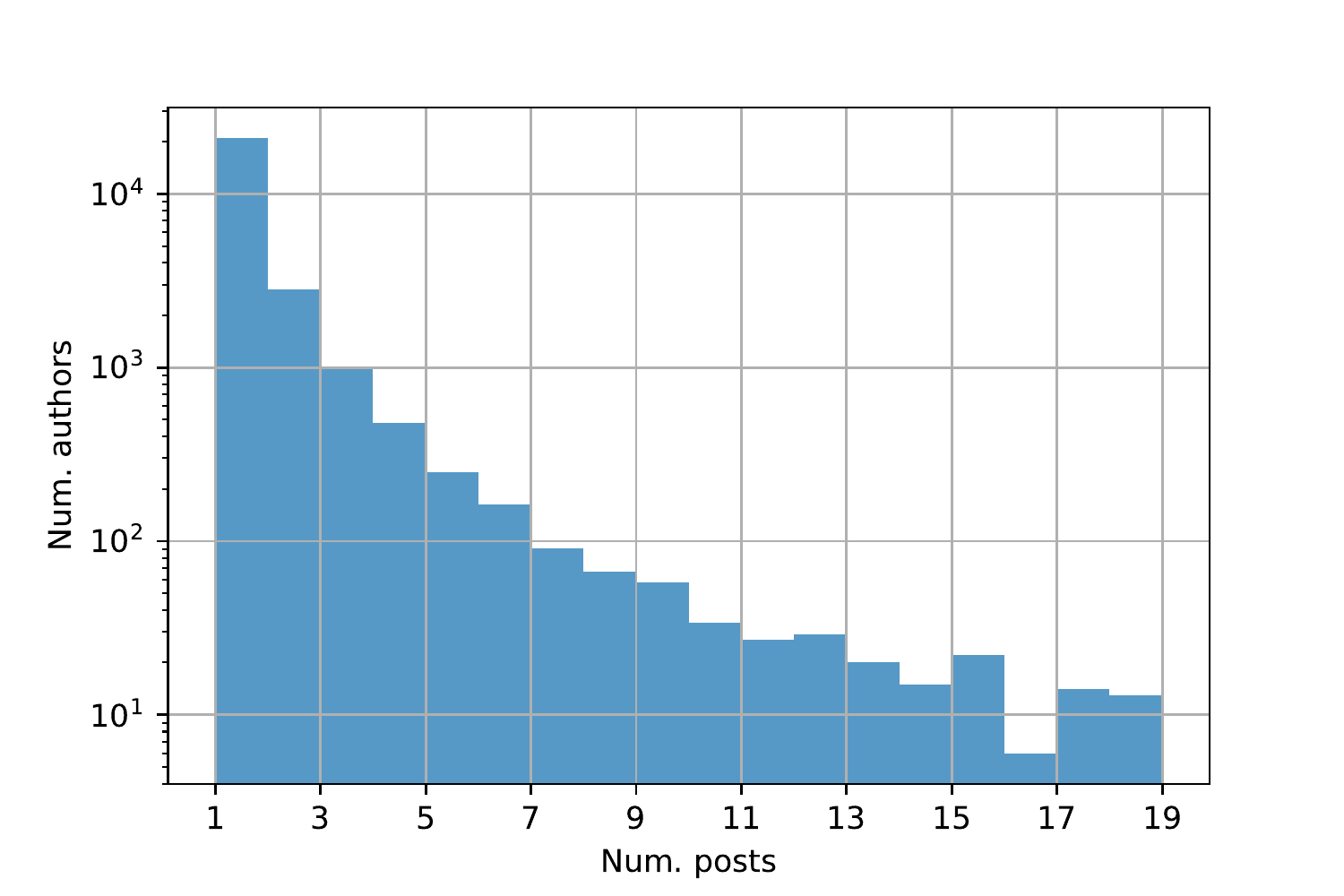}
    \caption{Distribution of the number of posts per author in our set of sociopolitical ChangeMyView posts.}
    \label{fig:dist-num-posts}
  \end{subfigure}
  \hspace{1cm}
  \begin{subfigure}{0.45\linewidth}
    \centering
    \includegraphics[width=0.9\linewidth]{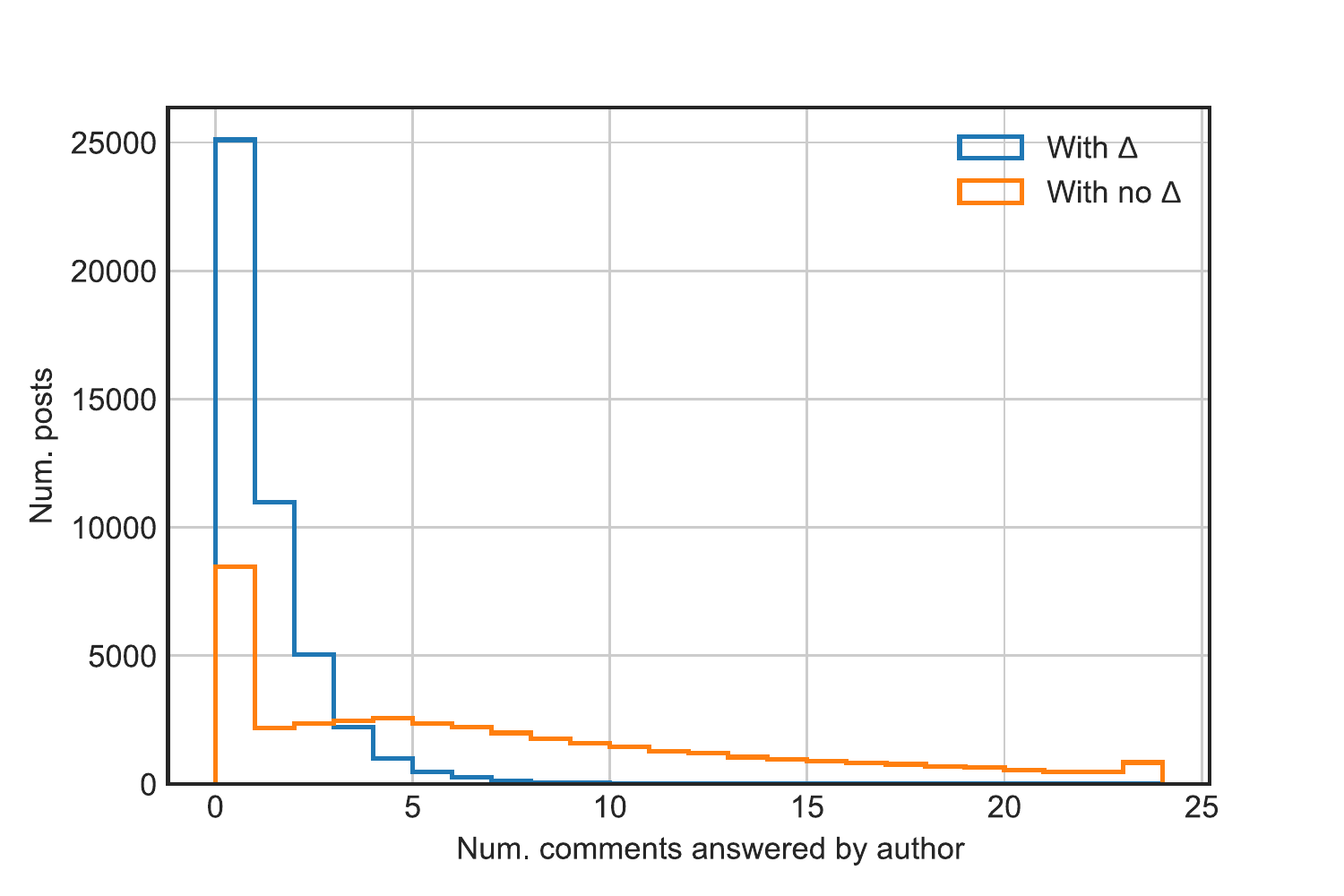}
    \caption{Distribution of the number of posts with a given number of comments for each category (with \cmvdelta and with no \cmvdelta), in our set of sociopolitical \mbox{ChangeMyView} posts.}
    \label{fig:dist-posts-delta}
  \end{subfigure}  
  \caption{Descriptive statistics of the sociopolitical posts and comments in \rcmv.}
  \label{fig:dists-all}
\end{figure}

We apply the classifier in order to find all \rcmv posts with sociopolitical topic. Here, we further characterize the \rcmv data set gathered this way. Figure~\ref{fig:sociopol-cmv-over-time} shows the number of posts per month over the time span of nine years considered in our analysis. Unlike the number of posts, which fluctuates over time, the fraction of posts that are sociopolitical is rather stable, suggesting that the discussion is not dominated by any event in particular. Figure~\ref{fig:dist-num-posts} reports the distribution of the number of posts per author in this set.

Finally, in our analysis we focus on the distinction between comments that received a \cmvdelta from the original author---which indicates the author admits to have changed their view after reading such comment---and those comments that did not receive one. Of the \num{3690687} comments, \num{38165} were awarded a \cmvdelta. As an additional control group, we find \num{504550} comments that did receive an answer from the original poster, and therefore received their attention, but did not obtain a \cmvdelta: this is further evidence that the original author did not consider such comments view-changing.
We report the distribution of the number of posts answered by the original author with a given number of comments in each of these two categories (with and without \cmvdelta) per post in Figure~\ref{fig:dist-posts-delta} (posts that have comments both with \cmvdelta and without count towards both distributions).

\section*{Social dimensions}

\begin{figure*}[]
    \centering
    \includegraphics[width=0.99\linewidth]{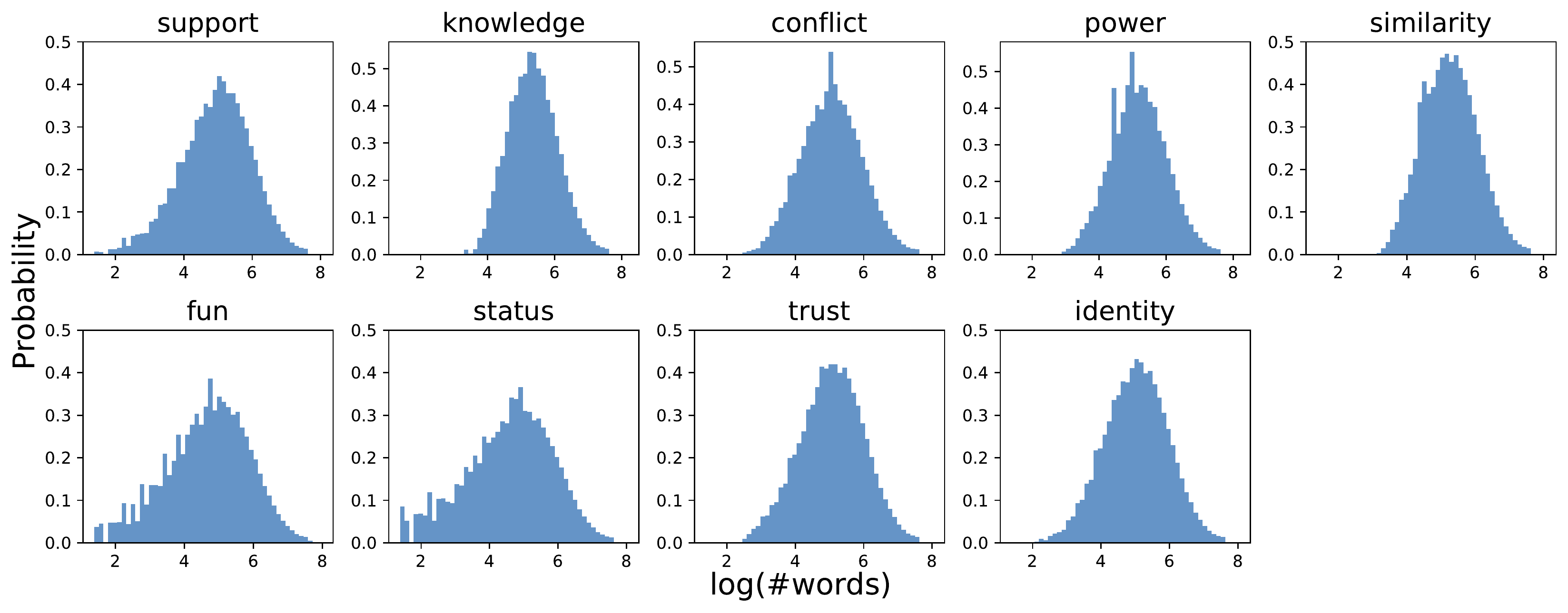}
    \caption{Distribution of (log) number of words in comments with dimension $d$.}
    \label{fig:word_length_log_distribution}
\end{figure*}

Figure~\ref{fig:word_length_log_distribution} shows the probability distribution of comment length across dimensions. The typical length of messages may vary considerably across dimensions; for example, comments conveying status tend to be much shorter that knowledge-exchange comments.

\begin{figure*}[]
    \centering
    \includegraphics[width=0.99\linewidth]{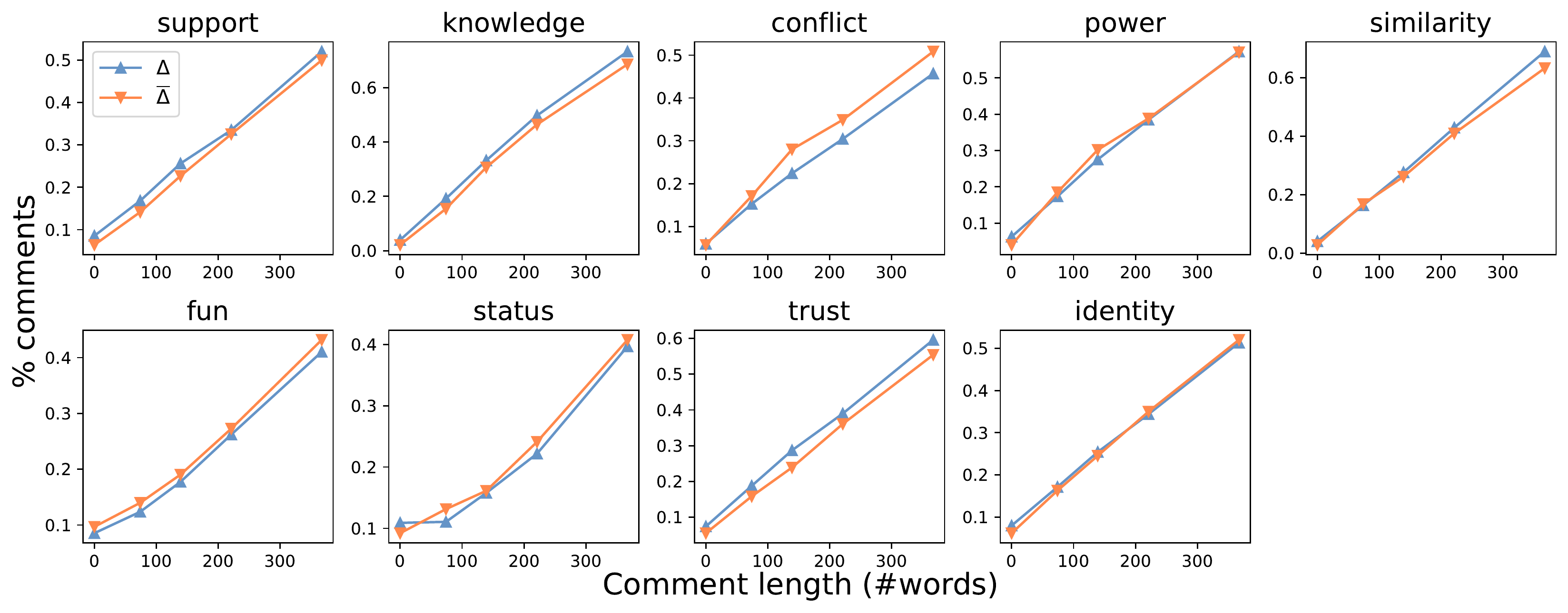}
    \caption{Fraction of comments of a given length that have dimension $d$, separately for comments with and without \cmvdelta.}
    \label{fig:dimensions_vs_length}
\end{figure*}

Figure~\ref{fig:dimensions_vs_length} shows the fraction of comments with dimension $d$ among all the comments with a given range of length. We consider five length classes, corresponding to the quintiles of the length distribution. The probability of finding a dimension in a comment increases linearly with the comment length.

\begin{figure*}[]
    \centering
    \includegraphics[width=0.49\linewidth]{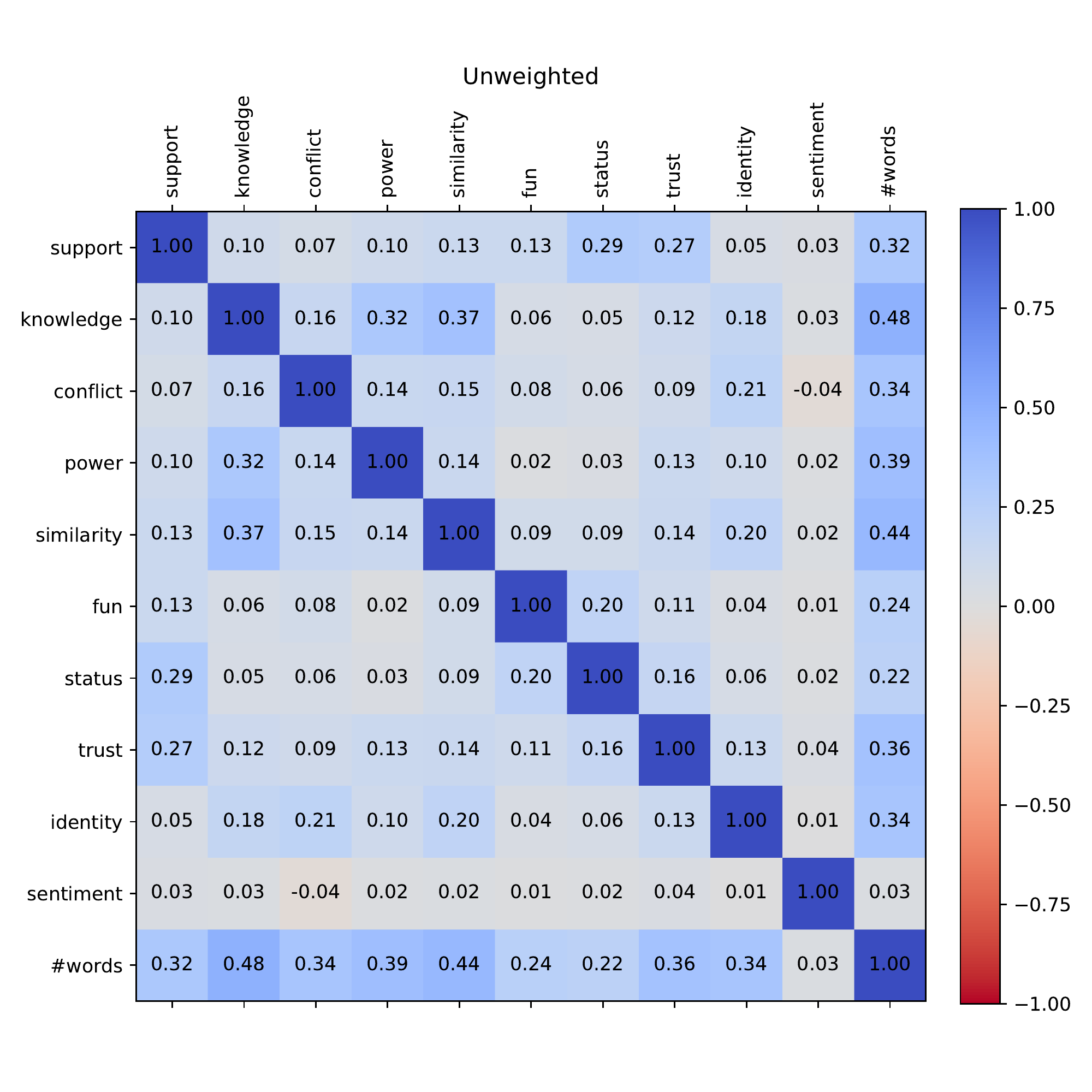}
    \includegraphics[width=0.49\linewidth]{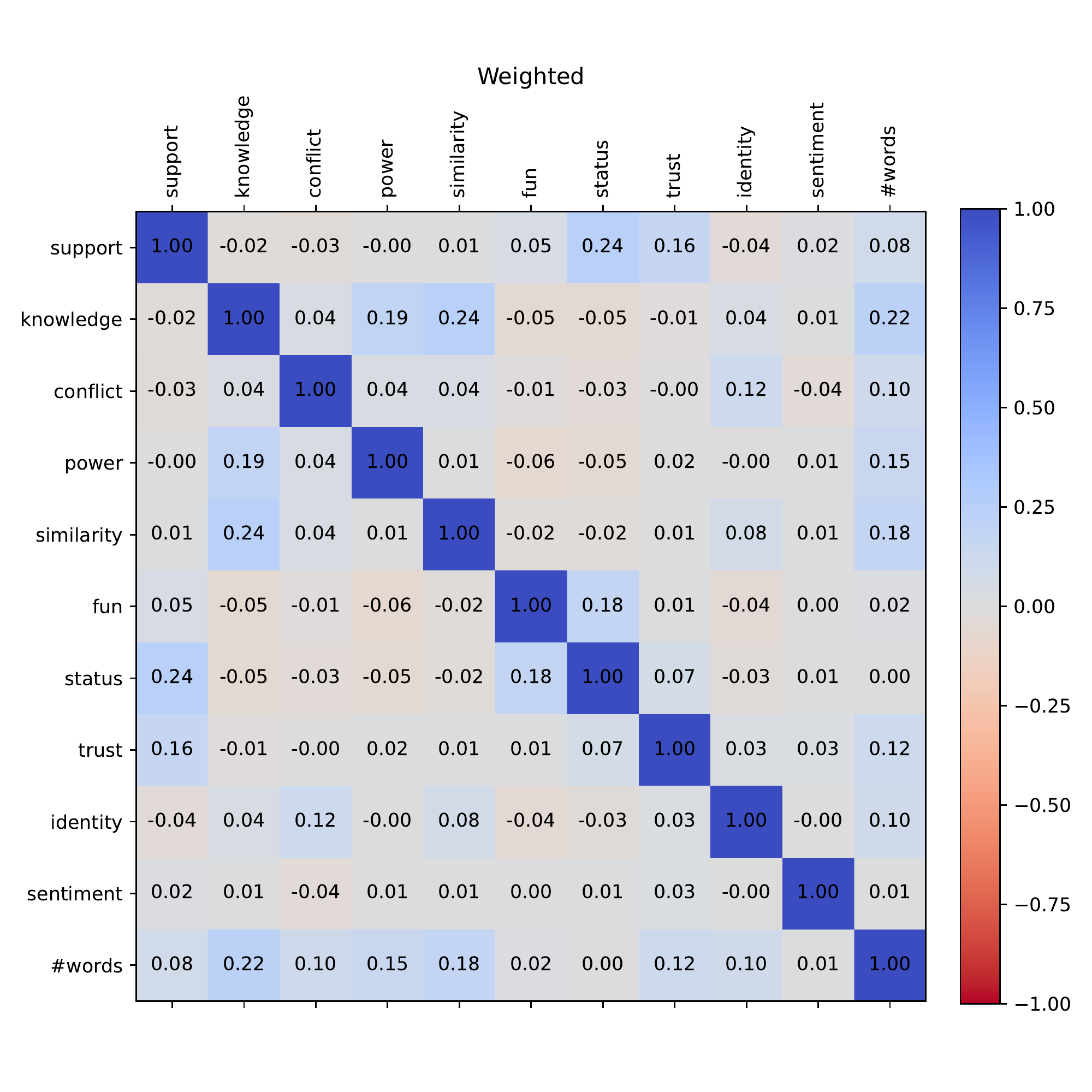}
    \caption{Correlation matrix between variables before (left) and after (right) weight discounting.}
    \label{fig:corr_matrix}
\end{figure*}

Figure~\ref{fig:corr_matrix} shows the cross-correlations between the dimension scores, for all the dimension pairs (plus sentiment scores and comment length, measured in number of words). On the left, we report correlations computed on the original values $s_d(m)$. On the right, we report correlations computed on the weight-discounted values $d(m)$. The weight-discounting reduces the cross-correlations considerably.

\section*{Opinion change}

Table~\ref{tab:comments-social-dim-log-reg} shows the results of a logistic regression to predict whether a comment got a $\Delta$ by using ($i$) all the dimensions including fun, and ($ii$) with sentiment scores in addition to all the dimensions.
The social dimension of \emph{Fun} is not significant, we therefore remove it from the other regression models.

Table~\ref{tab:homophily-log-reg-length} repeats the same analysis from Table~\ref{tab:homophily-log-reg} but including information about the length of the message.
We do so by including as an independent variable the quantity $Z(\log l)$, where $l$ is the length of the message and $Z$ is a Z-score standardization.
We apply a logarithmic scaling since the length of a message is broadly distributed.
Note that this regression model is spurious, since there is an interdependence between the length of a message and the social dimensions conveyed by it. Namely, it usually takes more words to express some social dimensions rather than others. For example, to express \emph{knowledge}, one typically requires to use a relatively large number of words to articulate an argument, whereas \emph{status} can be conveyed effectively with just a few words of admiration.

Nonetheless, these results show that many of our findings are robust: in particular, all of the dimensions are highly significant (as in Table~\ref{tab:homophily-log-reg}).

\begin{figure*}[t]
    \centering
    \begin{tabular}{ccc}
      \raisebox{-0.43\height}{\includegraphics[angle=-90,origin=c,width=0.25\linewidth]{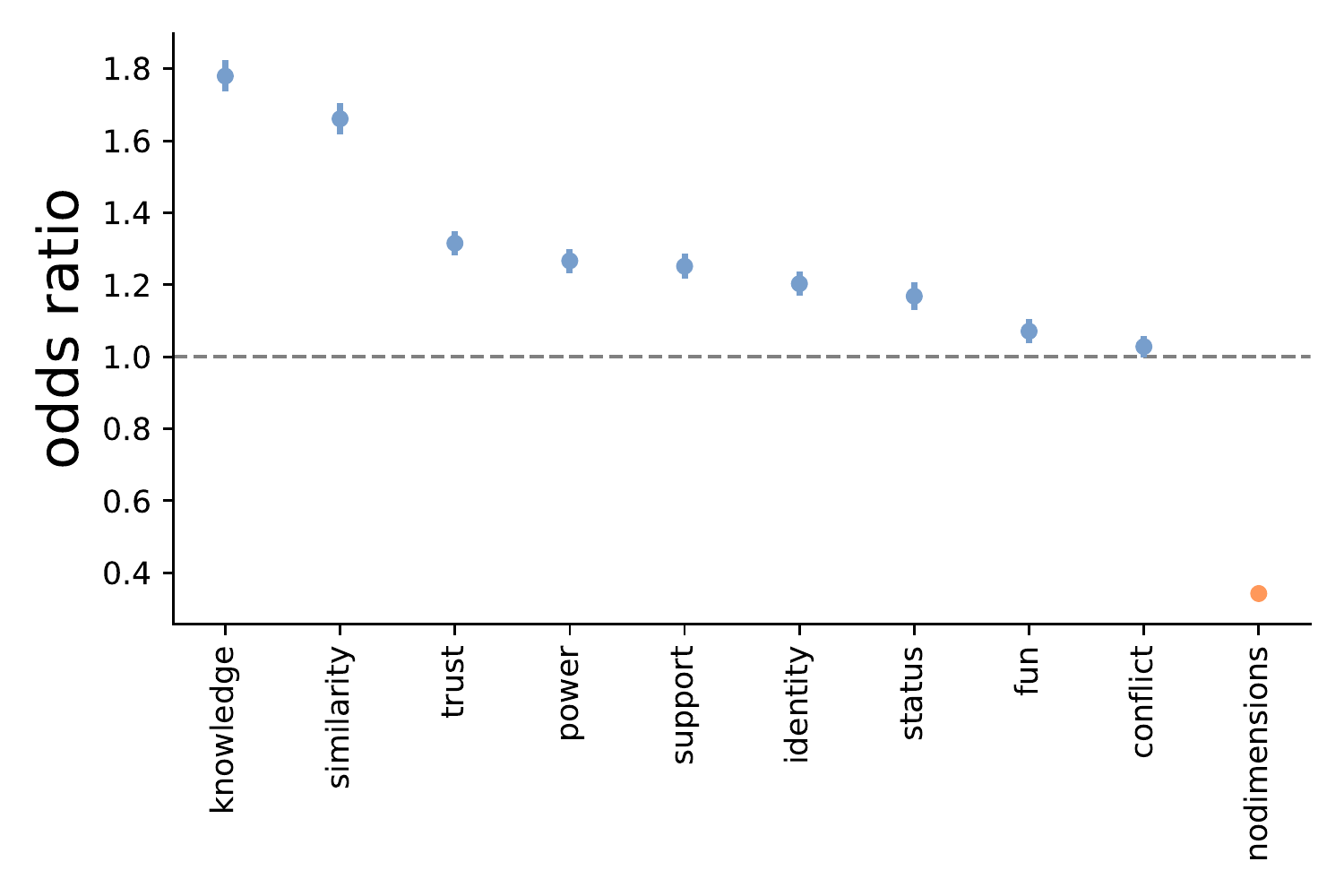}} &
      \raisebox{-0.43\height}{\includegraphics[angle=-90,origin=c,width=0.25\linewidth]{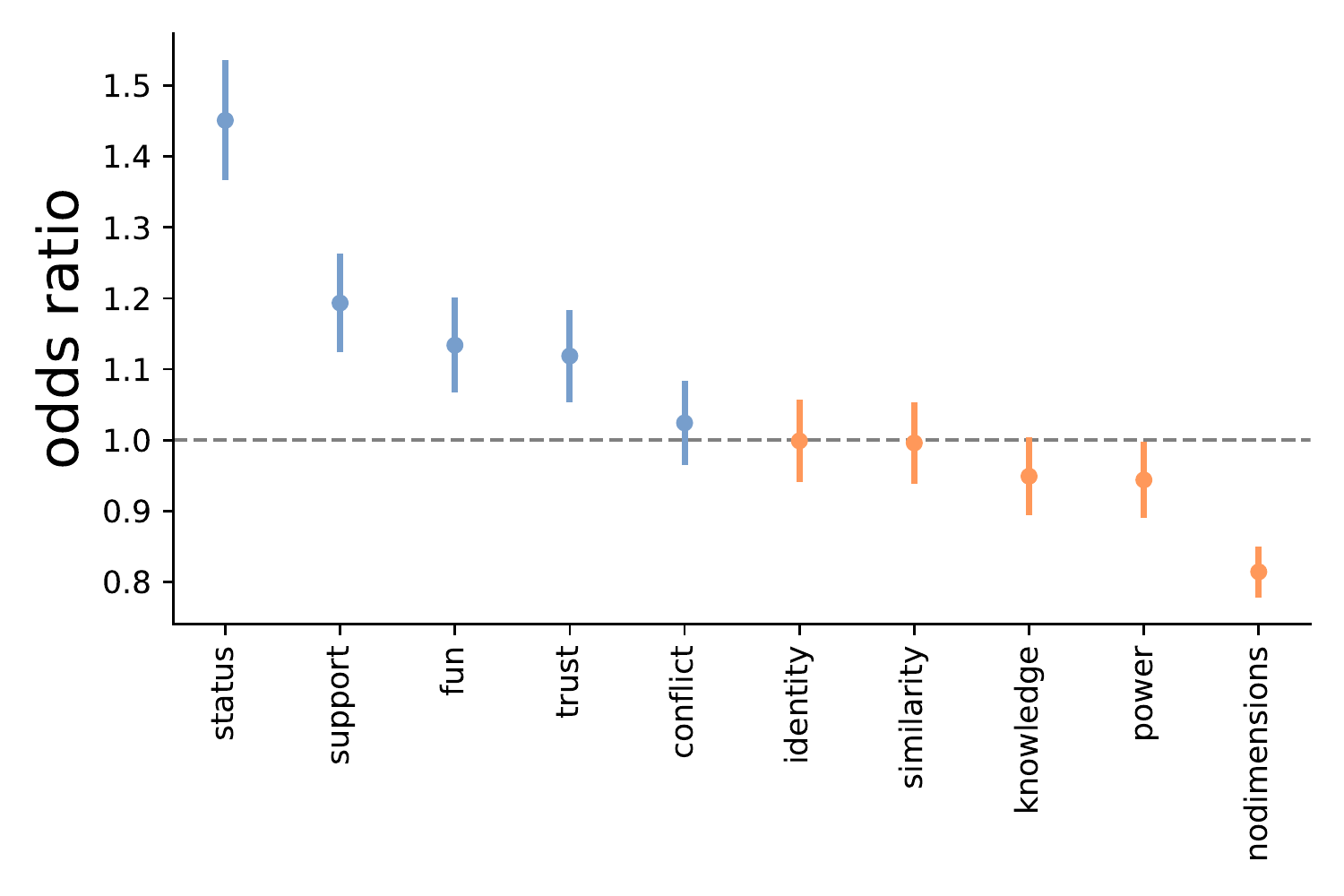}} &
      \raisebox{-0.5\height}{\includegraphics[width=0.45\linewidth]{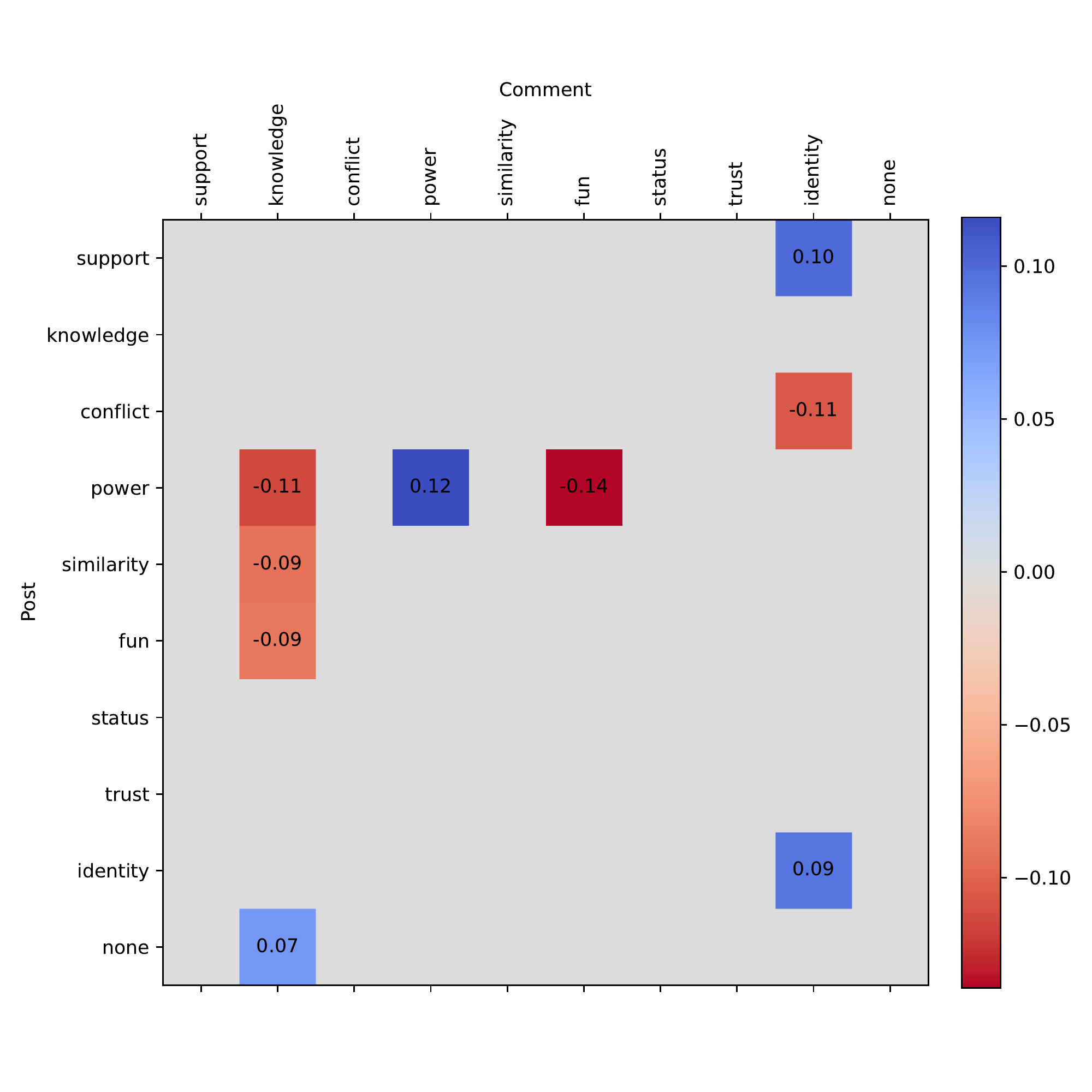}} \\
      (a) & (b) & (c) \\
    \end{tabular}
    \caption{Odd ratios of containing a dimension in opinion-changing messages versus the others in the case of (a) comments and (b) posts. Here, we considered only comments with an answer from the original poster. On the right (c), we report only the statistically significant odds ratios ($p < 0.01$) for interactions between dimensions in comments and posts. Results are qualitatively similar to Figure~\ref{fig:odd-ratios}.}
    \label{fig:odd-ratios-with-answers}
\end{figure*}

Figure~\ref{fig:odd-ratios-with-answers} shows the odds ratios calculated considering only comments that got a reply from the OP. Figure~\ref{fig:odd-ratios-with-answers}a shows the the odds ratios of a social dimension being conveyed by comments with $\Delta$ compared to comments with no $\Delta$. Figure~\ref{fig:odd-ratios-with-answers}b shows the the odds ratios of a social dimension being conveyed by posts for which a$\Delta$ was awarded, compared to posts whose authors did not give any $\Delta$. Figure Figure~\ref{fig:odd-ratios-with-answers}c shows a matrix of interaction between the intent of the poster and that of the commenter. The results are qualitatively very similar to those obtained when considering all the comments, including those that got no reply from the OP.

Tables~\ref{table:balanced},~\ref{table:randomized}, and~\ref{table:higher-sociopol-threshold}, report results of experiments we conducted as robustness checks.

Table~\ref{table:balanced} reports the results of the regression model fit on a balanced data set obtained by undersampling negative examples. The results are very similar to the those reported in Table~2 in the main manuscript. The predictive power is slightly improved, simply because we remove the added difficulty of class imbalance and additional noise provided by the negative examples.

Table~\ref{table:randomized} reports the results of a regression model fit on a randomized dataset that we obtained by randomly shuffling the social dimensions associated to each example, so that the association between opinion change and social dimensions is disrupted. In this experiment, no social dimension appears as significant, as one would expect.

Last, Table~\ref{table:higher-sociopol-threshold} reports the results of the regression that we obtain when considering a more conservative threshold of the sociopolitical classifier (i.e., moving the threshold from 0.50 to 0.75).
Such a change does not affect any of the significance levels that we record in the main regression model---most coefficients do not vary by more than $0.01$ compared to those presented in Table 2 in the main manuscript. Also the variation in the Pseudo-$R^2$ is minimal. We can therefore conclude that, even if the accuracy of our sociopolitical classifier is not perfect, our results seem robust to minor classification errors.

\begin{table}[]
    \centering
    \caption{
    Odds ratios obtained by a logistic regression that considers all social dimensions (including the non-significant \emph{fun}), with and without traditional sentiment scores.
    We indicate with asterisks the statistically significant correlations (with one, two, or three asterisks corresponding to $p < 0.05$, $p < 0.01$ and $p < 0.001$ respectively).
    P-values are corrected according to the Benjamini-Hochberg procedure, to reduce the chance of spurious correlation emerging because of the high number of factors we consider.
    }
    \small
\begin{tabular}{lll}
\toprule
Adj. Pseudo-$R^2$ &   0.01520 &   0.01643 \\ \midrule
Intercept         &  0.010*** &  0.010*** \\
Support           &  1.111*** &  1.096*** \\
Knowledge         &  1.226*** &  1.217*** \\
Conflict          &  1.029*** &  1.024*** \\
Power             &  1.099*** &  1.097*** \\
Similarity        &  1.118*** &  1.110*** \\
Fun               &     0.999 &     0.993 \\
Status            &  0.935*** &  0.931*** \\
Trust             &  1.155*** &  1.143*** \\
Identity          &  1.088*** &  1.085*** \\
Sentiment Neg.    &           &  1.056*** \\
Sentiment Pos.    &           &  1.111*** \\
\bottomrule
\end{tabular}

    \label{tab:comments-social-dim-log-reg}
\end{table}

\begin{table}[]
    \centering
    \caption{
    Odds ratios obtained by logistic regression when considering also the length of the comment.
    Each column corresponds to a model with a specific set of variables.
    A description of each variable is given in Table~\ref{tab:homophily-log-reg-explaination}.
    We indicate with asterisks the statistically significant correlations (with one, two, or three asterisks corresponding to $p < 0.05$, $p < 0.01$ and $p < 0.001$ respectively).
    P-values are corrected according to the Benjamini-Hochberg procedure, to reduce the chance of spurious correlation emerging because of the high number of factors we consider.
    }
    \small
\begin{tabular}{lllllllll}
\toprule
{} &         A &         B &         C &         D &         E &         F &         G &         H \\
\midrule
Adj. Pseudo-$R^2$            &   0.06898 &   0.07054 &   0.07479 &   0.07283 &   0.07042 &   0.07193 &   0.07615 &   0.07421 \\ \midrule \midrule
Intercept                    &  0.007*** &  0.007*** &  0.007*** &  0.007*** &  0.007*** &  0.007*** &  0.007*** &  0.007*** \\ \addlinespace
Support                      &  \dotfill &  \dotfill &  \dotfill &  \dotfill &  1.033*** &  1.032*** &  1.034*** &  1.033*** \\
Knowledge                    &  \dotfill &  \dotfill &  \dotfill &  \dotfill &  1.066*** &  1.065*** &  1.066*** &  1.065*** \\
Conflict                     &  \dotfill &  \dotfill &  \dotfill &  \dotfill &  0.925*** &  0.927*** &  0.926*** &  0.926*** \\
Power                        &  \dotfill &  \dotfill &  \dotfill &  \dotfill &  0.974*** &  0.976*** &  0.973*** &  0.974*** \\
Similarity                   &  \dotfill &  \dotfill &  \dotfill &  \dotfill &    0.988* &   0.986** &    0.989* &    0.988* \\
Status                       &  \dotfill &  \dotfill &  \dotfill &  \dotfill &  0.916*** &  0.915*** &  0.923*** &  0.920*** \\
Trust                        &  \dotfill &  \dotfill &  \dotfill &  \dotfill &  1.051*** &  1.050*** &  1.050*** &  1.050*** \\
Identity                     &  \dotfill &  \dotfill &  \dotfill &  \dotfill &     0.999 &     0.999 &     0.999 &     0.999 \\ \addlinespace
Sentiment Pos.               &  1.054*** &  1.052*** &  1.056*** &  1.054*** &  1.043*** &  1.042*** &  1.044*** &  1.042*** \\
Sentiment Neg.               &  0.979*** &   0.981** &  0.980*** &  0.980*** &     0.996 &     0.998 &     0.997 &     0.997 \\ \addlinespace
Comment Left-Wing            &  \dotfill &     0.984 &  \dotfill &  \dotfill &  \dotfill &     0.983 &  \dotfill &  \dotfill \\
Comment Right-Wing           &  \dotfill &  0.772*** &  \dotfill &  \dotfill &  \dotfill &  0.776*** &  \dotfill &  \dotfill \\
Post Left-Wing               &  \dotfill &  0.878*** &  \dotfill &  \dotfill &  \dotfill &  0.880*** &  \dotfill &  \dotfill \\
Post Right-Wing              &  \dotfill &  0.847*** &  \dotfill &  \dotfill &  \dotfill &  0.846*** &  \dotfill &  \dotfill \\ \addlinespace
Both Polarized               &  \dotfill &  \dotfill &  \dotfill &  0.332*** &  \dotfill &  \dotfill &  \dotfill &  0.334*** \\
Both Polarized \& Diff. Side &  \dotfill &  \dotfill &  \dotfill &  2.345*** &  \dotfill &  \dotfill &  \dotfill &  2.338*** \\
Diff. Side                   &  \dotfill &  \dotfill &  \dotfill &     1.017 &  \dotfill &  \dotfill &  \dotfill &     1.017 \\
Shared Group                 &  \dotfill &  \dotfill &  0.297*** &  \dotfill &  \dotfill &  \dotfill &  0.299*** &  \dotfill \\
Log Length                   &  2.929*** &  2.929*** &  2.905*** &  2.914*** &  2.904*** &  2.906*** &  2.879*** &  2.890*** \\
\bottomrule
\end{tabular}

    \label{tab:homophily-log-reg-length}
\end{table}

\begin{table}
\caption{Odds ratios obtained by logistic regression with a balanced data set.}
\label{table:balanced}
\centering
\footnotesize
\begin{tabular}{lllllllll}
\toprule
{} &         A &         B &         C &         D &         E &         F &         G &         H \\
\midrule
Adj. Pseudo-$R^2$            &   0.00658 &   0.00938 &   0.01555 &   0.02031 &   0.03804 &   0.04065 &   0.04644 &   0.05093 \\ \midrule \midrule
Intercept                    &   0.981** &  1.064*** &  1.030*** &  1.053*** &  0.902*** &    0.978* &  0.946*** &  0.967*** \\ \addlinespace
Support                      &  \dotfill &  \dotfill &  \dotfill &  \dotfill &  1.126*** &  1.125*** &  1.128*** &  1.128*** \\
Knowledge                    &  \dotfill &  \dotfill &  \dotfill &  \dotfill &  1.272*** &  1.270*** &  1.269*** &  1.269*** \\
Conflict                     &  \dotfill &  \dotfill &  \dotfill &  \dotfill &  1.036*** &  1.038*** &  1.037*** &  1.036*** \\
Power                        &  \dotfill &  \dotfill &  \dotfill &  \dotfill &  1.138*** &  1.141*** &  1.139*** &  1.136*** \\
Similarity                   &  \dotfill &  \dotfill &  \dotfill &  \dotfill &  1.152*** &  1.149*** &  1.151*** &  1.152*** \\
Status                       &  \dotfill &  \dotfill &  \dotfill &  \dotfill &  0.920*** &  0.919*** &  0.924*** &  0.927*** \\
Trust                        &  \dotfill &  \dotfill &  \dotfill &  \dotfill &  1.177*** &  1.177*** &  1.177*** &  1.178*** \\
Identity                     &  \dotfill &  \dotfill &  \dotfill &  \dotfill &  1.102*** &  1.103*** &  1.104*** &  1.104*** \\ \addlinespace
Sentiment Pos.               &  1.211*** &  1.208*** &  1.211*** &  1.214*** &  1.136*** &  1.134*** &  1.135*** &  1.138*** \\
Sentiment Neg.               &  1.095*** &  1.097*** &  1.097*** &  1.096*** &  1.071*** &  1.072*** &  1.072*** &  1.072*** \\ \addlinespace
Comment Left-Wing            &  \dotfill &     1.016 &  \dotfill &  \dotfill &  \dotfill &     1.013 &  \dotfill &  \dotfill \\
Comment Right-Wing           &  \dotfill &  0.788*** &  \dotfill &  \dotfill &  \dotfill &  0.788*** &  \dotfill &  \dotfill \\
Post Left-Wing               &  \dotfill &  0.884*** &  \dotfill &  \dotfill &  \dotfill &  0.890*** &  \dotfill &  \dotfill \\
Post Right-Wing              &  \dotfill &  0.835*** &  \dotfill &  \dotfill &  \dotfill &  0.836*** &  \dotfill &  \dotfill \\ \addlinespace
Both Polarized               &  \dotfill &  \dotfill &  0.329*** &  \dotfill &  \dotfill &  \dotfill &  0.332*** &  \dotfill \\
Both Polarized \& Diff. Side &  \dotfill &  \dotfill &  2.565*** &  \dotfill &  \dotfill &  \dotfill &  2.540*** &  \dotfill \\
Diff. Side                   &  \dotfill &  \dotfill &     1.021 &  \dotfill &  \dotfill &  \dotfill &     1.025 &  \dotfill \\
Shared Group                 &  \dotfill &  \dotfill &  \dotfill &  0.294*** &  \dotfill &  \dotfill &  \dotfill &  0.297*** \\
\bottomrule
\end{tabular}
\end{table}

\begin{table}
\caption{Odds ratios obtained by logistic regression with randomized social dimensions.}
\label{table:randomized}
\centering
\footnotesize
\begin{tabular}{lllllllll}
\toprule
{} &         A &         B &         C &         D &         E &         F &         G &         H \\
\midrule
Adj. Pseudo-$R^2$            &   0.00305 &   0.00446 &   0.00720 &   0.00935 &   0.00303 &   0.00444 &   0.00718 &   0.00933 \\ \midrule \midrule
Intercept                    &  0.010*** &  0.011*** &  0.011*** &  0.011*** &  0.010*** &  0.011*** &  0.011*** &  0.011*** \\ \addlinespace
Support                      &  \dotfill &  \dotfill &  \dotfill &  \dotfill &     1.005 &     1.005 &     1.005 &     1.005 \\
Knowledge                    &  \dotfill &  \dotfill &  \dotfill &  \dotfill &     1.006 &     1.006 &     1.006 &     1.006 \\
Conflict                     &  \dotfill &  \dotfill &  \dotfill &  \dotfill &     1.002 &     1.002 &     1.002 &     1.002 \\
Power                        &  \dotfill &  \dotfill &  \dotfill &  \dotfill &     0.997 &     0.997 &     0.997 &     0.997 \\
Similarity                   &  \dotfill &  \dotfill &  \dotfill &  \dotfill &     0.993 &     0.993 &     0.993 &     0.993 \\
Status                       &  \dotfill &  \dotfill &  \dotfill &  \dotfill &     1.008 &     1.008 &     1.008 &     1.008 \\
Trust                        &  \dotfill &  \dotfill &  \dotfill &  \dotfill &     0.996 &     0.996 &     0.996 &     0.996 \\
Identity                     &  \dotfill &  \dotfill &  \dotfill &  \dotfill &     1.005 &     1.005 &     1.005 &     1.005 \\ \addlinespace
Sentiment Pos.               &  1.174*** &  1.173*** &  1.175*** &  1.177*** &  1.174*** &  1.173*** &  1.175*** &  1.177*** \\
Sentiment Neg.               &  1.080*** &  1.082*** &  1.081*** &  1.081*** &  1.080*** &  1.082*** &  1.081*** &  1.081*** \\ \addlinespace
Comment Left-Wing            &  \dotfill &     1.014 &  \dotfill &  \dotfill &  \dotfill &     1.014 &  \dotfill &  \dotfill \\
Comment Right-Wing           &  \dotfill &  0.790*** &  \dotfill &  \dotfill &  \dotfill &  0.790*** &  \dotfill &  \dotfill \\
Post Left-Wing               &  \dotfill &  0.867*** &  \dotfill &  \dotfill &  \dotfill &  0.867*** &  \dotfill &  \dotfill \\
Post Right-Wing              &  \dotfill &  0.852*** &  \dotfill &  \dotfill &  \dotfill &  0.852*** &  \dotfill &  \dotfill \\ \addlinespace
Both Polarized               &  \dotfill &  \dotfill &  0.321*** &  \dotfill &  \dotfill &  \dotfill &  0.321*** &  \dotfill \\
Both Polarized \& Diff. Side &  \dotfill &  \dotfill &  2.555*** &  \dotfill &  \dotfill &  \dotfill &  2.555*** &  \dotfill \\
Diff. Side                   &  \dotfill &  \dotfill &     1.022 &  \dotfill &  \dotfill &  \dotfill &     1.022 &  \dotfill \\
Shared Group                 &  \dotfill &  \dotfill &  \dotfill &  0.286*** &  \dotfill &  \dotfill &  \dotfill &  0.286*** \\
\bottomrule
\end{tabular}

\end{table}

\begin{table}
\caption{Odds ratios obtained by logistic regression, analogously to Table~2 in the manuscript, but with a more precise sociopolitical classifier (threshold of $0.75$ instead of 0.5). }
\label{table:higher-sociopol-threshold}
\centering
\footnotesize
\begin{tabular}{lllllllll}
\toprule
{} &         A &         B &         C &         D &         E &         F &         G &         H \\
\midrule
Adj. Pseudo-$R^2$            &   0.00285 &   0.00418 &   0.00695 &   0.00925 &   0.01642 &   0.01770 &   0.02043 &   0.02274 \\ \midrule \midrule
Intercept                    &  0.010*** &  0.011*** &  0.010*** &  0.011*** &  0.009*** &  0.010*** &  0.010*** &  0.010*** \\ \addlinespace
Support                      &  \dotfill &  \dotfill &  \dotfill &  \dotfill &  1.090*** &  1.089*** &  1.091*** &  1.092*** \\
Knowledge                    &  \dotfill &  \dotfill &  \dotfill &  \dotfill &  1.214*** &  1.213*** &  1.213*** &  1.213*** \\
Conflict                     &  \dotfill &  \dotfill &  \dotfill &  \dotfill &  1.030*** &  1.032*** &  1.031*** &  1.031*** \\
Power                        &  \dotfill &  \dotfill &  \dotfill &  \dotfill &  1.104*** &  1.106*** &  1.104*** &  1.103*** \\
Similarity                   &  \dotfill &  \dotfill &  \dotfill &  \dotfill &  1.109*** &  1.107*** &  1.109*** &  1.111*** \\
Status                       &  \dotfill &  \dotfill &  \dotfill &  \dotfill &  0.928*** &  0.928*** &  0.933*** &  0.936*** \\
Trust                        &  \dotfill &  \dotfill &  \dotfill &  \dotfill &  1.147*** &  1.147*** &  1.147*** &  1.148*** \\
Identity                     &  \dotfill &  \dotfill &  \dotfill &  \dotfill &  1.091*** &  1.091*** &  1.092*** &  1.092*** \\ \addlinespace
Sentiment Pos.               &  1.170*** &  1.169*** &  1.171*** &  1.173*** &  1.107*** &  1.107*** &  1.107*** &  1.109*** \\
Sentiment Neg.               &  1.077*** &  1.080*** &  1.079*** &  1.079*** &  1.052*** &  1.054*** &  1.054*** &  1.053*** \\ \addlinespace
Comment Left-Wing            &  \dotfill &     1.012 &  \dotfill &  \dotfill &  \dotfill &     1.003 &  \dotfill &  \dotfill \\
Comment Right-Wing           &  \dotfill &  0.785*** &  \dotfill &  \dotfill &  \dotfill &  0.790*** &  \dotfill &  \dotfill \\
Post Left-Wing               &  \dotfill &  0.877*** &  \dotfill &  \dotfill &  \dotfill &  0.883*** &  \dotfill &  \dotfill \\
Post Right-Wing              &  \dotfill &  0.870*** &  \dotfill &  \dotfill &  \dotfill &  0.868*** &  \dotfill &  \dotfill \\ \addlinespace
Both Polarized               &  \dotfill &  \dotfill &  0.336*** &  \dotfill &  \dotfill &  \dotfill &  0.340*** &  \dotfill \\
Both Polarized \& Diff. Side &  \dotfill &  \dotfill &  2.372*** &  \dotfill &  \dotfill &  \dotfill &  2.333*** &  \dotfill \\
Diff. Side                   &  \dotfill &  \dotfill &  1.052*** &  \dotfill &  \dotfill &  \dotfill &  1.053*** &  \dotfill \\
Shared Group                 &  \dotfill &  \dotfill &  \dotfill &  0.288*** &  \dotfill &  \dotfill &  \dotfill &  0.289*** \\
\bottomrule
\end{tabular}
\end{table}

\paragraph{Homophily and opinion change.}
\label{sec:homophily}

\begin{figure}
  \begin{subfigure}{0.9\linewidth}
      \centering
      \includegraphics[width=0.49\linewidth]{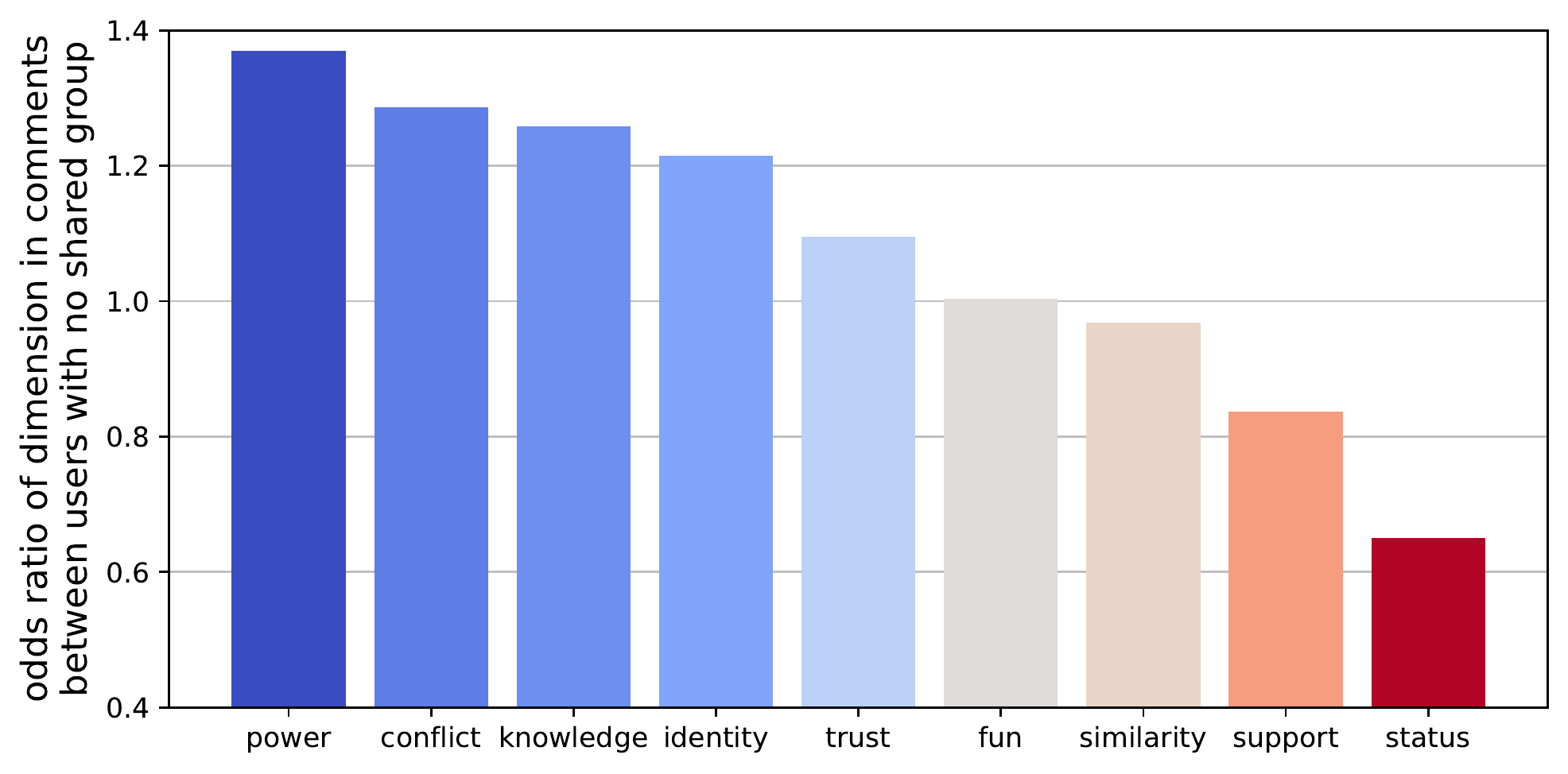}
      \includegraphics[width=0.49\linewidth]{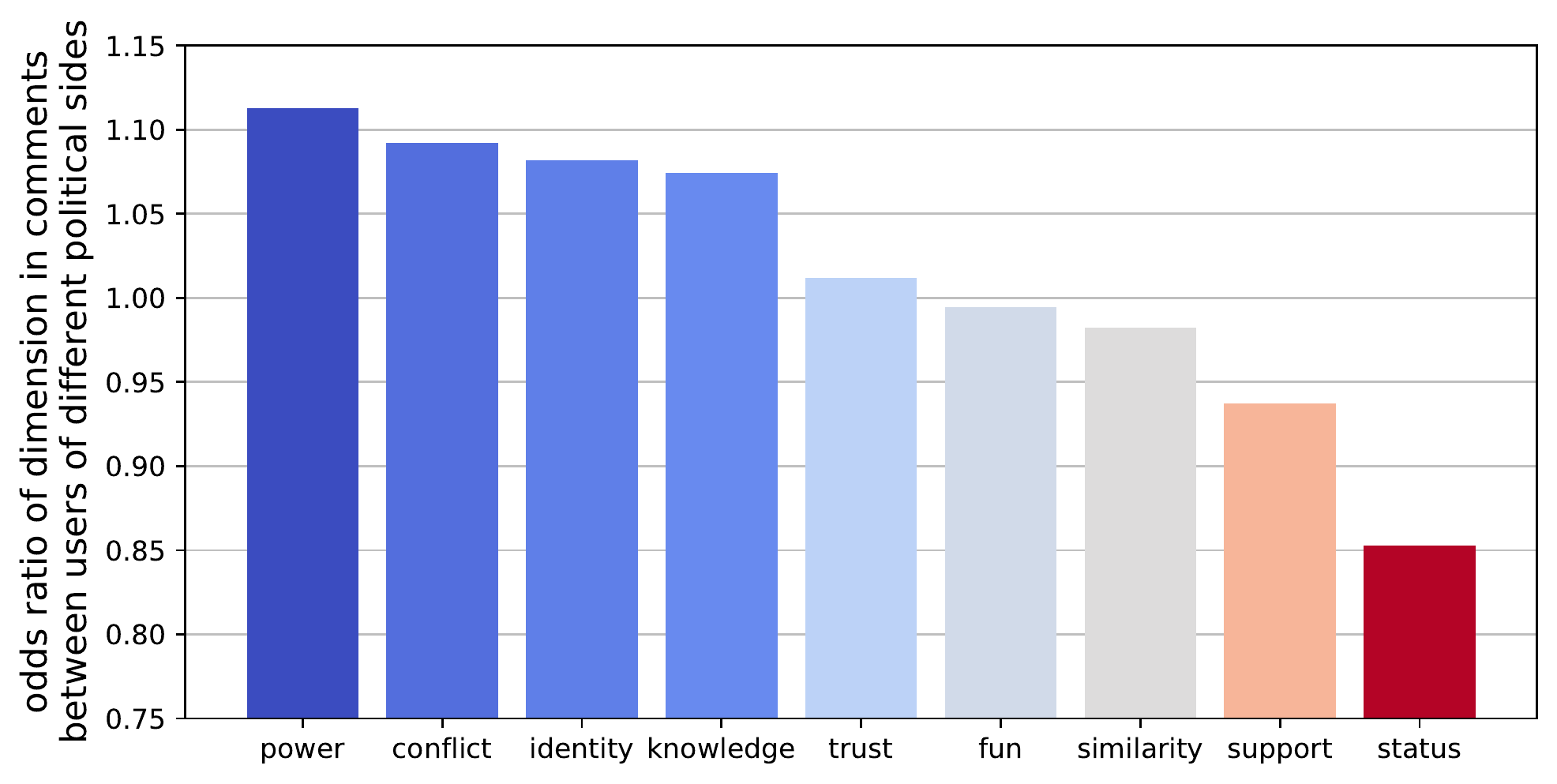}
      \caption{Odd ratios of presence of each dimension in comments between people with (left) no shared subreddits, (right) different political alignments.}
      \label{fig:profile_similarity}
  \end{subfigure}
  \\
  \begin{subfigure}{0.9\linewidth}
      \centering
      \includegraphics[trim={0 3cm 0 4cm},clip,width=0.45\linewidth]{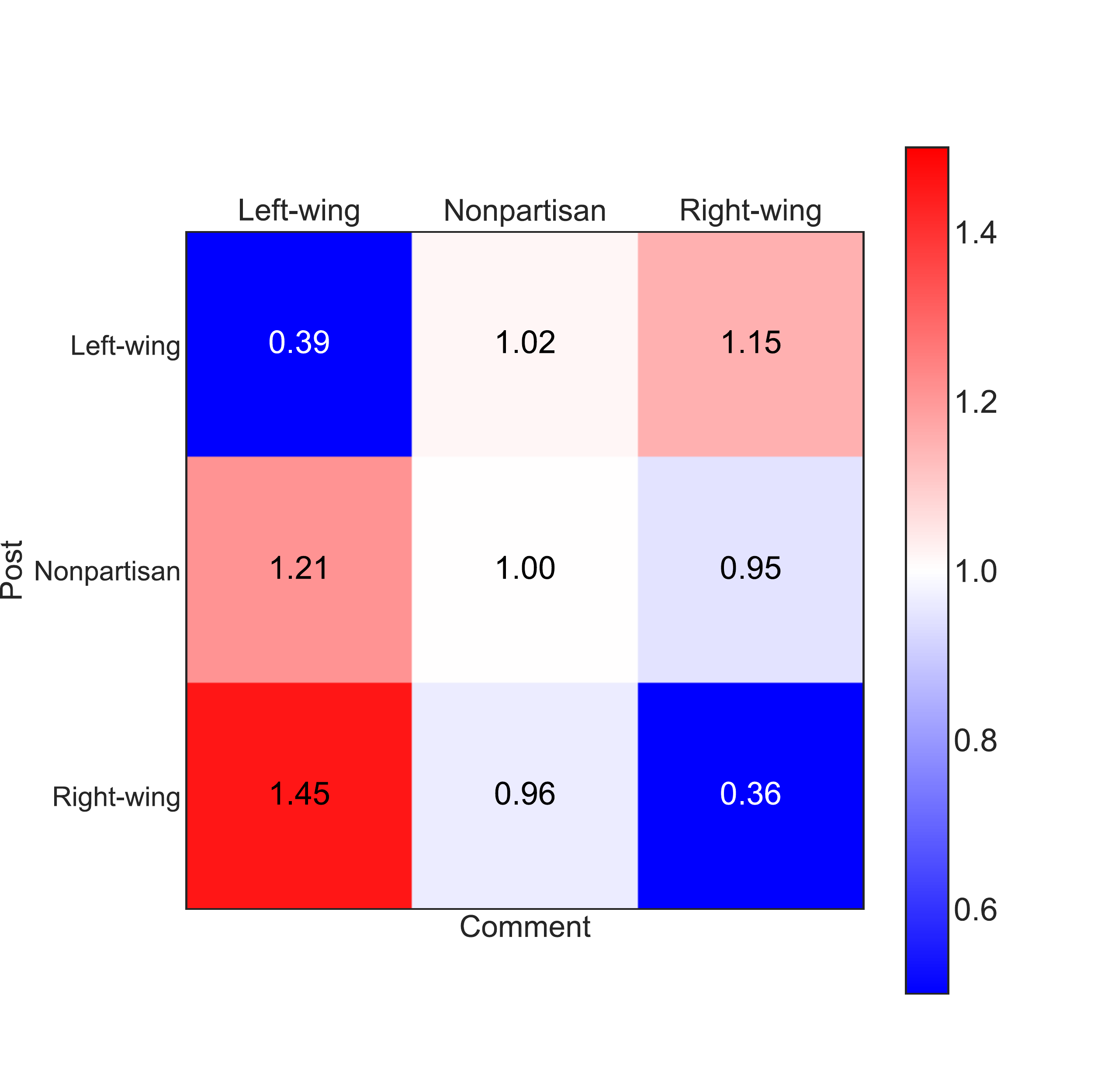}
      \caption{Odds ratio of presence of a $\Delta$ in comments between people with a given alignment combination, as estimated by a logistic regression model including these factors and the social dimensions. We normalize the odds ratio with respect to the interaction between two individuals with no detected political side.}
      \label{fig:homophily-matrix}
  \end{subfigure}
  \caption{Analysis of political alignment, in relation to (a)~social dimensions and (b)~opinion change.}
  \label{fig:homophily-matrix-all}
\end{figure}

Figure~\ref{fig:profile_similarity} shows the likelihood that different dimensions are present in comments that are exchanged by two users with different profiles. Figure~\ref{fig:profile_similarity} (left) shows the odds ratios of a dimension being present in a comment when the commenter and the author of the commented post participated in some political subreddits, but do not have one in common. Figure~\ref{fig:profile_similarity} (right) shows the odds rations of a dimension being present in a comment when the commenter and the poster belong to different ideologies (left-wing vs. right-wing). Comments flowing between users with different profiles are more likely to convey power and conflict and less likely to contain support or status. Also, knowledge-rich discussion happens between people who have different stances.

Figure~\ref{fig:homophily-matrix} shows the result of a logistic regression model that includes as independent variables the social dimensions of the comment, the political side of the commenter, and the political side of the poster. The dependent variable is, as in Table~\ref{tab:homophily-log-reg}, whether the comment is marked with a $\Delta$ or not (that is, if it is labelled as opinion-changing). We aggregate the resulting coefficients in order to obtain odds ratios for all of the nine possible configurations of these variables. Then, we normalize such odds ratios with respect to the interaction between two individual with no detected political side. From these results, we can observe:
\begin{enumerate}[]
  \item heterophily: individuals are more likely to change their opinion when confronted by an individual of the opposite political side;
  \item asymmetry between left-wing and right-wing individuals: a left-wing individual is more likely to receive a $\Delta$ from a right-wing one than vice versa.
\end{enumerate}

